\documentclass[12pt]{iopart}
\usepackage{iopams}
\usepackage{cite}
\usepackage{array}
\usepackage[dvipsnames]{xcolor}
\usepackage{graphicx}
\usepackage{hyperref}

\expandafter\let\csname equation*\endcsname\relax

\expandafter\let\csname endequation*\endcsname\relax

\usepackage{mathtools}
\eqnobysec

\begin{document}

\title[Hadamard renormalization for a charged scalar field]{Hadamard renormalization for a charged scalar field}

\author{Visakan Balakumar}

\address{Consortium for Fundamental Physics, School of Mathematics and Statistics,\\ Hicks Building, Hounsfield Road, Sheffield. S3 7RH United Kingdom}
\ead{VBalakumar1@sheffield.ac.uk}

\author{Elizabeth Winstanley}
\address{Consortium for Fundamental Physics, School of Mathematics and Statistics,\\ Hicks Building, Hounsfield Road, Sheffield. S3 7RH United Kingdom}
\ead{E.Winstanley@sheffield.ac.uk}

\begin{abstract}
The Hadamard representation of the Green's function of a quantum field on a curved space-time is a powerful tool for computations of renormalized expectation values.
We study the Hadamard form of the Feynman Green's function for a massive charged complex scalar field in an arbitrary number of space-time dimensions.
Explicit expressions for the coefficients in the Hadamard parametrix are given for two, three and four space-time dimensions. 
We then develop the formalism for the Hadamard renormalization of the expectation values of the scalar field condensate, current and stress-energy tensor. 
These results will have applications in the computation of renormalized expectation values for a charged quantum scalar field on a charged black hole space-time, and hence in addressing issues such as the quantum stability of the inner horizon. 
\end{abstract}


\vspace{2pc}

\section{Introduction}
\label{sec:intro}

The quest for a theory of quantum gravity, in which both space-time and matter are quantized, has yet to yield a definitive solution.
In the absence of such a theory, a semiclassical approach, namely quantum field theory on curved space-time (QFTCS), represents a first step.
In QFTCS, the space-time metric is treated classically, and the properties of quantum fields propagating on this classical background are studied.
Any successful theory of quantum gravity must reproduce the results of QFTCS in an appropriate limit, and therefore QFTCS is a nontrivial testing-ground for theories of quantum gravity.
QFTCS has also resulted in many deep results in its own right, such as the discovery of Hawking radiation \cite{Hawking:1974sw}, the Unruh effect  \cite{Unruh:1976db} 
and the creation of particles in an expanding universe \cite{Parker:1968mv,Parker:1969au,Parker:1971pt} (see also \cite{Crispino:2007eb,Birrell:1982ix,Fulling:1989nb,Wald:1995,Parker:2009uva} for reviews). 

The renormalized expectation value of the stress-energy tensor (RSET) $\langle {\hat {T}}_{\mu \nu } \rangle _{\rm {ren}}$ plays a central role in QFTCS.  Via the semiclassical Einstein equations
\begin{equation}
    G_{\mu \nu }+ \Lambda g_{\mu \nu } =  \langle {\hat {T}}_{\mu \nu } \rangle _{\rm {ren}}
    \label{eq:SCEE}
\end{equation}
(where $G_{\mu \nu }$ is the Einstein tensor, $\Lambda $ the cosmological constant, $g_{\mu \nu }$ the space-time metric and we are using units in which $8\pi G=c=\hbar =1$) 
the RSET governs how the quantum field will affect the space-time geometry.
However, since the stress-energy tensor operator involves products of field operators evaluated at the same space-time point, it is divergent and a naive computation of its expectation value will give an infinite result. 
It is therefore necessary to employ some kind of regularization and renormalization prescription.

There are many possible approaches to renormalizing the RSET (see, for example, \cite{Birrell:1982ix,Fulling:1989nb,Parker:2009uva}).
Amongst these, the axiomatic approach developed by Wald \cite{Wald:1995,Wald:1977up} has proved to be extremely powerful.
The stress-energy tensor operator is regularized by point-splitting \cite{DeWitt:1975ys,Christensen:1976vb,Christensen:1978yd}, considering the field operators acting at two distinct, but closely separated, space-time points.
The divergences in the RSET arise in the limit in which the space-time points are brought together.
These divergences are purely geometric and independent of the state of the quantum field. 
The RSET is computed by subtracting appropriate geometric divergent terms and then taking the coincidence limit of the space-time points.
Wald \cite{Wald:1995,Wald:1977up} has given a list of physically-motivated axioms which must be satisfied by the resulting RSET, and which determine the RSET up to the addition of a local conserved tensor. 
Such a renormalization ambiguity is to be expected in the absence of a full theory of quantum gravity and corresponds to the freedom to move any local conserved tensor from the left-hand-side to the right-hand-side of (\ref{eq:SCEE}). 

The RSET can be computed by applying a second-order differential operator to the Feynman Green's function $G_{\rm {F}}(x,x')$ of the quantum field.
The Feynman Green's function depends on two space-time points $x$ and $x'$, and is itself divergent in the limit $x'\rightarrow x$.
The divergences in the RSET can therefore be identified from the corresponding divergences in $G_{\rm {F}}(x,x')$.
In the original formulation of point-splitting regularization \cite{DeWitt:1975ys,Christensen:1976vb,Christensen:1978yd}, the parametrix giving the divergent terms in $G_{\rm {F}}(x,x')$ was constructed using a DeWitt-Schwinger expansion.
This is a special case of the Hadamard representation of $G_{\rm {F}}(x,x')$
\cite{Wald:1978pj}.
Wald showed that subtracting the divergent parts of the Hadamard parametrix from $G_{\rm {F}}(x,x')$, applying the appropriate second-order differential operator and taking the coincidence limit yields an RSET which satisfies his axioms. 
It has been rigorously established that the Hadamard approach yields valid results for the renormalized stress-energy tensor, which are unique apart from the anticipated freedom to add a local conserved tensor (see, for example, 
\cite{Moretti:1998rf,Moretti:1999ez,Moretti:1998rs,Moretti:1999fb,Hollands:2001nf,Hollands:2001fb,Moretti:2001qh,Hollands:2002ux,Hollands:2004yh} and 
\cite{Decanini:2005eg} for a more comprehensive list of papers on Hadamard renormalization). 

Hadamard renormalization following this approach has proved to be an  elegant method for computing the RSET and other expectation values.
The Hadamard prescription has been developed in detail for a massive neutral scalar field with arbitrary coupling to the curvature in any number of space-time dimensions \cite{Decanini:2005eg}.
Hadamard renormalization has also been applied to the electromagnetic field \cite{Brown:1986tj},
St\"uckelberg massive electromagnetic field \cite{Belokogne:2015etf},
one-loop quantum gravity \cite{Allen:1987bn},
$p$-forms \cite{Folacci:1990eb} and fermions \cite{Najmi:1985yi,Hollands:1999fc,Dappiaggi:2009xj,Lewis:2019iwu}.

In this paper we extend the Hadamard formalism to a massive charged complex scalar field on a curved space-time with an arbitrary number of dimensions.
Our primary motivation is to develop the machinery necessary for the computation of the RSET for a charged quantum scalar field on a charged Reissner-Nordstr\"om black hole background.
Unlike the uncharged Schwarzschild black hole, a Reissner-Nordstr\"om black hole possesses an inner (Cauchy) horizon as well as an event horizon.
The inner horizon is classically unstable, with a weak singularity forming as a result of the backreaction of classical perturbations  \cite{Simpson:1973ua,Gursel:1979zza,Matzner:1979zz,Gursel:1979zz,Hiscock:1981,Poisson:1990eh,Ori:1991zz,Brady:1995ni,Burko:1997zy,Hod:1998gy}.
It is expected that there will also be a quantum instability at the inner horizon, with a stronger singularity forming as a result of divergences in quantum expectation values \cite{Hiscock:1977qe,Birrell:1978th,Hiscock:1980wr,Novikov:1980ni}.
The precise nature of quantum effects on the classical instability remains an open question.
Recently this has begun to be studied in detail, as new techniques have been developed for computing renormalized expectation values inside the event horizon of a black hole \cite{Lanir:2017oia,Lanir:2018rap}. 
An analysis of the leading-order asymptotics of the RSET for a massless, minimally coupled quantum scalar field  \cite{Sela:2018xko} in both the Hartle-Hawking \cite{Hartle:1976tp} and Unruh \cite{Unruh:1976db} states
found that the divergence  at the inner horizon is weaker than expected.
Numerical computations of the vacuum polarization (the expectation value of the square of the quantum scalar field) \cite{Lanir:2018vgb} reveal good agreement with this asymptotic analysis.
In the Hartle-Hawking and Unruh states, two of the components of the RSET have recently been computed on the inner horizon \cite{Zilberman:2019buh}, indicating that a curvature singularity forms at the inner horizon due to the back-reaction of the neutral scalar field.
At the same time, it has been shown that there exists a quantum state for which the RSET is regular at the inner horizon \cite{Taylor:2019qqc} so the back-reaction in this case would not lead to the formation of a singularity.
All this recent work considers only a neutral scalar field, whereas it is the presence of the classical electromagnetic field which leads to the inner horizon in the first place.
Therefore, computations of the RSET for a charged scalar field may help to shed further light onto this question.

This paper represents a first step in this direction, since the geometric divergent terms in the Hadamard parametrix (which we derive in this paper) must be known before the RSET can be computed.
In our analysis we make no assumptions about the background space-time geometry or the fixed, classical, electromagnetic field with which the charged scalar field interacts.
Therefore our results are applicable, not only to black hole space-times, but also to scalar QED on cosmological space-times, for which the RSET has recently been computed using adiabatic regularization in two  \cite{Ferreiro:2018qzr} and four \cite{Ferreiro:2018qdi} space-time dimensions.

The renormalization of the stress-energy tensor for a charged quantum scalar field has been previously considered in four space-time dimensions within the framework of DeWitt-Schwinger regularization in early work by Boulware \cite{Boulware:1978hy} and subsequently by Herman and Hiscock \cite{Herman:1995hm}. 
Generalizing the methodology of \cite{Christensen:1976vb,Christensen:1978yd}, Herman and Hiscock derive
 the quartic, quadratic, linear, logarithmic and finite renormalization counterterms required for the computation of the RSET,
 together with the linear and finite renormalization counterterms
 needed in computing the renormalized expectation value of the current operator $\langle {\hat {J}}^{\mu } \rangle _{\rm {ren}}$.
 This latter expectation value governs the back-reaction of the charged quantum scalar field on the classical electromagnetic field via the semiclassical Maxwell equations
 \begin{equation}
     \nabla _{\mu } F^{\mu \nu } = 4\pi \langle {\hat {J}}^{\nu } \rangle _{\rm {ren}}
     \label{eq:SCME}
 \end{equation}
 where $F^{\mu \nu }$ is the electromagnetic field, $\nabla _{\mu }$ the space-time covariant derivative and we are using Gaussian units.
 The analysis in \cite{Herman:1995hm} uses the Hadamard function $G^{{\rm {(1)}}}(x,x')=\langle {\hat {\Phi }}(x) {\hat {\Phi }}^{*}(x') + {\hat {\Phi }}^{*}(x'){\hat {\Phi }}(x)  \rangle $  where a star ${}^{*}$ denotes the complex conjugate.
 The Hadamard function $G^{{\rm {(1)}}}(x,x')$ is related to the Feynman propagator $G_{\rm {F}}(x,x')$ by the relation  $G_{\rm {F}}(x,x')={\overline {G}}(x,x') - \frac{1}{2}iG^{{\rm {(1)}}}(x,x')$, where ${\overline {G}}(x,x')$ is one half the sum of the advanced and retarded Green's functions.
 We emphasise that the Hadamard elementary function $G^{\rm {(1)}}(x,x')$ is {\it {not}} the same as the Hadamard representation of the Feynman Green's function $G_{\rm {F}}(x,x')$, which is our focus in this paper.
A method for computing the imaginary part of $G^{\rm {(1)}}(x,x')$ in a Euclideanized static, spherically symmetric space-time is developed in \cite{Herman:1998dz}.
More recently, the Seeley-DeWitt coefficients for a charged scalar field appearing in the DeWitt-Schwinger expansion have been studied in an alternative approach using heat kernel methods \cite{Ivanov:2019prj,Ivanov:2019uxu}.
 
 The outline of this paper is as follows.
 In Section \ref{sec:scalar}, we give the Hadamard parametrix of the Feynman Green's function for a massive complex charged scalar field with arbitrary coupling to the curvature and in arbitrary space-time dimensions, in terms of sesquisymmetric biscalars $U(x,x')$, $V(x,x')$ and $W(x,x')$, which have power series expansions in the square of the geodesic distance between the space-time points.
 The recurrence relations satisfied by the expansion coefficients (which are themselves biscalars) are derived in Section \ref{sec:Hadamard}, and solved explicitly for two, three and four space-time dimensions, up to the order required for the computation of the RSET.
 Expressions for the renormalized expectation values of the current and RSET are found in Section \ref{sec:expvals} in terms of quantities which depend on the quantum state of the field.
 We compare our results with those arising from other approaches to regularization in Section~\ref{sec:comparison}.
 Section \ref{sec:conc} contains our conclusions and discussion.

\section{Feynman Green's function for a charged scalar field}
\label{sec:scalar}

In $d$ space-time dimensions, we consider a massive, charged, complex scalar field $\Phi $ of mass $m$ and charge $q$ satisfying the equation
\begin{equation}
    \left[ D_{\mu }D^{\mu } - m^{2}-\xi R \right] \Phi =0
    \label{eq:scalar}
\end{equation}
where $D_{\mu }=\nabla _{\mu } - {\rm {i}}qA_{\mu }$ is the covariant derivative, with $A_{\mu }$ the electromagnetic potential; $R$ is the Ricci scalar curvature and $\xi $ a coupling constant. 
Throughout this paper the metric has signature $(-,+,\ldots , +)$.
If $ \xi =0$ the scalar field is minimally coupled to the Ricci scalar curvature and, in $d$ space-time dimensions, the scalar field is conformally coupled to the curvature if $\xi = \xi _{{\rm {c}}}$, where
\begin{equation}
    \xi _{{\rm{c}}} = \frac{d - 2}{4\left( d-1 \right) } . 
    \label{eq:xic}
\end{equation}

We assume that the scalar field has been quantized but that the electromagnetic field and space-time metric remain classical. 
The scalar field is assumed to be in a Hadamard state (this is a reasonable assumption for physical states \cite{Fewster:2013lqa}). 
We consider the Feynman Green's function $G_{{\rm {F}}}(x,x')$ for this state, 
which is given by the expectation value of the  time-ordered product
\begin{equation}
    -{\rm {i}}G_{\rm {F}}(x,x')= \left\langle T\left[ {\hat {\Phi }}(x) {\hat {\Phi }}^{\dagger }(x') \right] \right\rangle 
\end{equation}
where ${\hat {\Phi }}^{\dagger }$ denotes the adjoint field operator (which is not equal to ${\hat {\Phi }}$ for a complex scalar field).
The Feynman Green's function $G_{\rm {F}}(x,x')$
is a biscalar function of the distinct space-time points $x$ and $x'$ and satisfies the inhomogeneous scalar field equation
\begin{equation}
   \left[ D_{\mu }D^{\mu } - m^{2}-\xi R \right] G_{{\rm {F}}}(x,x') =
   - \left[ -g(x) \right] ^{-\frac{1}{2}} \delta ^{d} (x-x')
   \label{eq:inhom}
\end{equation}
where $g(x)$ is the determinant of the space-time metric and $\delta ^{d}(x-x')$ is the $d$-dimensional Dirac delta function.
We assume henceforth that the space-time point $x'$ lies within a normal neighbourhood of the point $x$, so that there is a unique geodesic connecting the two points.

Our assumption that the quantum state of the scalar field is Hadamard dictates the form of the Feynman propagator $G_{\rm {F}}(x,x')$ for closely separated points, depending on the number of space-time dimensions.
In all cases the Hadamard expansion of $G_{\rm {F}}(x,x')$ depends on the geodetic interval $\sigma (x,x')$, which is one-half the square of the geodesic distance between $x$ and $x'$, and satisfies the equation
\begin{equation}
    2\sigma = g_{\mu \nu }\sigma ^{;\mu }\sigma ^{;\nu }
    \label{eq:sigma}
\end{equation}
where we use a semicolon $;$ to denote a space-time covariant derivative, that is $\sigma ^{;\mu } = \nabla ^{\mu }\sigma $.
Depending on the number of space-time dimensions, the Feynman propagator will depend on two or three biscalars, which we denote $U(x,x')$, $V(x,x')$ and $W(x,x')$ in accordance with standard notation \cite{Decanini:2005eg}, and which are regular in the limit $x'\rightarrow x$.
In the list below we give the form of the Feynman propagator in terms of $U(x,x')$, $V(x,x')$, $W(x,x')$ and $\sigma (x,x')$ for different numbers of space-time dimensions, and also the expansions of $U(x,x')$, $V(x,x')$ and $W(x,x')$ in terms of powers of $\sigma $. 
In each case we include a factor ${\rm {i}}\epsilon $ as $\epsilon \rightarrow 0_{+}$ to ensure that the singularity structure of $G_{\rm {F}}(x,x')$ is consistent with the definition of the Feynman propagator as a time-ordered product.  We use a superscript of the form $(d)$ to denote the number of space-time dimensions $d$. 
\begin{description}
\item[$d=2$]
In two space-time dimensions, the Hadamard form of the Feynman propagator is
\begin{equation}
  \hspace{-1cm}  -{\rm {i}}G_{\rm {F}}^{(2)}(x,x') = \alpha ^{(2)}  \left\{ 
    V^{(2)}(x,x') \ln \left[ \frac{\sigma (x,x')}{\ell ^{2}} + {\rm {i}}\epsilon \right]
    +W^{(2)}(x,x')
    \right\}  ,  
    \label{eq:GF_d2}
\end{equation}
where 
\begin{equation}
    \alpha ^{(2)} =  \frac{1}{4\pi } 
    \label{eq:alpha2}
\end{equation}
and the biscalars $V^{(2)}(x,x')$ and $W^{(2)}(x,x')$ can be written as power series in the geodetic interval $\sigma $:
\begin{subequations}
\label{eq:VWd2}
\begin{eqnarray}
    V^{(2)}(x,x') & = &  \sum _{n=0}^{\infty } V_{n}^{(2)} (x,x') \sigma ^{n}(x,x'), 
    \label{eq:Vd2} 
    \\
    W^{(2)}(x,x') & = &  \sum _{n=0}^{\infty } W_{n}^{(2)} (x,x') \sigma ^{n}(x,x').
    \label{eq:Wd2}
\end{eqnarray}
\end{subequations}
\item[$d=2p$, $p>1$]
When the number of space-time dimensions is even and greater than two, so that $d=2p$ with $p=2,3,\ldots $, the Hadamard expansion of the Feynman propagator takes the form
\begin{eqnarray}
    \hspace{-1cm}-{\rm {i}}G_{\rm {F}}^{(2p)}(x,x') & = &  \alpha ^{(2p)}
    \left\{ 
    \frac{U^{(2p)}(x,x')}{\left[ \sigma (x,x')+{\rm {i}}\epsilon  \right] ^{p-1}} 
    \right. 
    \nonumber \\ & & \left.
    +
    V^{(2p)}(x,x') \ln \left[ \frac{\sigma (x,x')}{\ell ^{2}} + {\rm {i}}\epsilon \right]
    + W^{(2p)}(x,x') 
    \right\} ,
    \label{eq:GF_deven}
\end{eqnarray}
where 
\begin{equation}
    \alpha ^{(2p)} = \frac{\Gamma \left( p-1 \right)}{2\left( 2\pi  \right) ^{p} }
\label{eq:alpha2p}
\end{equation}
and the expansions of $U^{(2p)}(x,x')$, $V^{(2p)}(x,x')$ and $W^{(2p)}(x,x')$ in powers of $\sigma (x,x')$ are now
\begin{subequations}
\label{eq:UVWdeven}
\begin{eqnarray}
   U^{(2p)}(x,x') &  = &  \sum _{n=0}^{p-2} U_{n}^{(2p)} (x,x') \sigma ^{n}(x,x'), 
   \label{eq:Udeven}
   \\
   V^{(2p)}(x,x') & = &  \sum _{n=0}^{\infty } V_{n}^{(2p)} (x,x') \sigma ^{n}(x,x'), 
   \label{eq:Vdeven}
   \\
    W^{(2p)}(x,x') &  = &  \sum _{n=0}^{\infty } W_{n}^{(2p)} (x,x') \sigma ^{n}(x,x').
    \label{eq:Wdeven}
\end{eqnarray}
\end{subequations}
\item[$d=2p+1$]
When the number of space-time dimensions is odd, $d=2p+1$ with $p=0,1,2,\ldots $, the Feynman propagator has the Hadamard form
\begin{equation}
   \hspace{-2cm} -{\rm {i}}G_{\rm {F}}^{(2p+1)}(x,x') = \alpha ^{(2p+1)}
    \left\{ 
    \frac{U^{(2p+1)}(x,x')}{\left[ \sigma (x,x')+{\rm {i}}\epsilon  \right] ^{p-\frac{1}{2}}} + W^{(2p+1)}(x,x') 
    \right\} ,
    \label{eq:GF_dodd}
\end{equation}
where
\begin{equation}
   \alpha ^{(2p+1)}  = \frac{\Gamma \left( p-\frac{1}{2} \right)}{2\left( 2\pi  \right) ^{p+\frac{1}{2}} }
\end{equation}
and
\begin{subequations}
\label{eq:UWdodd}
\begin{eqnarray}
   U^{(2p+1)}(x,x') & = &  \sum _{n=0}^{\infty } U_{n}^{(2p+1)} (x,x') \sigma ^{n}(x,x'), 
   \label{eq:Udodd}
    \\
    W^{(2p+1)}(x,x') & = &  \sum _{n=0}^{\infty } W_{n}^{(2p+1)} (x,x') \sigma ^{n}(x,x').
    \label{eq:Wdodd}
\end{eqnarray}
\end{subequations}
\end{description}
The quantity $\ell $ introduced in the Feynman Green's functions (\ref{eq:GF_d2}, \ref{eq:GF_deven}) for even space-time dimensions is an arbitrary length scale, required to ensure that the argument of the logarithm is dimensionless. 
Both the Hadamard parametrices (\ref{eq:GF_d2}, \ref{eq:GF_deven}, \ref{eq:GF_dodd}) and the general forms of the power series expansions of $U(x,x')$, $V(x,x')$ and $W(x,x')$ (\ref{eq:VWd2}, \ref{eq:UVWdeven}, \ref{eq:UWdodd}) are unchanged from those for a neutral scalar field \cite{Decanini:2005eg} because including the gauge potential $A_{\mu }$ does not alter the principal part of the PDE (\ref{eq:inhom}).

There is one significant difference in the Feynman Green's function $G^{(d)}_{{\rm {F}}}(x,x')$ and Hadamard parametrices for a charged complex scalar field compared to a real neutral scalar field considered in \cite{Decanini:2005eg}. 
For a real neutral scalar field, both the Feynman Green's function $G^{(d)}_{{\rm {F}}}(x,x')$ and the Hadamard coefficients $U^{(d)}(x,x')$, $V^{(d)}(x,x')$ and $W^{(d)}(x,x')$ are real symmetric biscalars.
However, for the complex charged scalar field, these will be complex sesquisymmetric biscalars \cite{Kaminski:2019adk},  satisfying the symmetry relation
\begin{equation}
    K(x,x')=K^{*}(x',x).
    \label{eq:seqsui}
\end{equation}
The condition (\ref{eq:seqsui}) will also be satisfied by the Hadamard expansion coefficients $U^{(d)}_{n}(x,x')$, $V^{(d)}_{n}(x,x')$ and $W^{(d)}_{n}(x,x')$.
Now consider a general symmetric sesquisymmetric biscalar $K(x,x')$ satisfying (\ref{eq:seqsui}), and suppose that this has a covariant Taylor series expansion of the form
\begin{equation}
    K(x,x')= k_{0}(x)+k_{1\mu }(x)\sigma ^{;\mu } + k_{2\mu \nu }(x)\sigma ^{;\mu }\sigma ^{;\nu } + k_{3\mu \nu \lambda }(x)\sigma ^{;\mu }\sigma ^{;\nu } \sigma ^{;\lambda }+ \ldots ,
\end{equation}
where the coefficients $k_{0}$, $k_{1\mu }$, $k_{2\mu \nu }$ and $k_{3\mu \nu \lambda }$ are complex and depend only on the space-time point $x$.
From (\ref{eq:seqsui}), the real part of $K(x,x')$ {\it {is}} a symmetric biscalar.
This means that $k_{0}(x)$ must be real and that the real parts of $k_{1\mu }$ and $k_{3\mu \nu \lambda }$ are fixed to be \cite{Brown:1983zy,Brown:1984gs,Bernard:1986vc}:
\begin{subequations}
\label{eq:rek1k3}
\begin{eqnarray}
\Re \left[ k_{1\mu } \right] & = & 
-\frac{1}{2} k_{0;\mu }
\label{eq:rek1}
 \\
\Re \left[ k_{3\mu \nu \lambda } \right] & = & -\frac{1}{2}\Re \left[ k_{2(\mu \nu ; \lambda )} \right] 
+ \frac{1}{24} k_{0;(\mu \nu \lambda )}.
\label{eq:rek3}
\end{eqnarray}
\end{subequations}
In addition, we find that the imaginary part of $k_{2\mu \nu }$ is also fixed by (\ref{eq:seqsui}):
\begin{equation}
    \Im \left[ k_{2\mu \nu } \right] = \frac{1}{4{\rm {i}}} \left[ k_{0;\mu \nu } + 2k^{*}_{1(\mu ; \nu )} \right] = -\frac{1}{2} \Im \left[ k_{1(\mu ;\nu )} \right] 
    \label{eq:imk2}
\end{equation}
where we have simplified using (\ref{eq:rek1k3}).
In (\ref{eq:rek1k3}, \ref{eq:imk2}), we have used the notation $\Re [k]$ to denote the real part and $\Im [k]$ the imaginary part of the quantity $k$.

\section{Hadamard coefficients}
\label{sec:Hadamard}

While the Hadamard expansion of the Feynman propagator is the same for a charged as for an uncharged scalar field, it is anticipated that the form of the Hadamard coefficients $U^{(d)}_{n}(x,x')$, $V^{(d)}_{n}(x,x')$ and $W^{(d)}_{n}(x,x')$ will be changed by the introduction of the electromagnetic potential $A_{\mu }$ into the scalar field equation (\ref{eq:inhom}). 
In this section we derive the recurrence relations satisfied by the Hadamard coefficients, and give their explicit forms for $d=2$, $3$ and $4$ to the order required for the renormalization of the stress-energy tensor.

To derive the recurrence relations, we make use of (\ref{eq:sigma}) for the geodetic distance. 
In addition, the recurrence relations involve the Van Vleck-Morette determinant $\Delta (x,x')$, a regular biscalar defined by 
\begin{equation}
    \Delta (x,x') = - \left[  -g(x)\right] ^{-\frac{1}{2}} \det 
    \left[ - \sigma _{;\mu \nu '}(x,x') \right] \left[  -g(x')\right] ^{-\frac{1}{2}}
\end{equation}
where a subscript $;\nu '$ denotes the space-time covariant derivative $\nabla _{\nu '}$ with respect to the space-time point $x'$.
The Van Vleck-Morette determinant is related to the D'Alembertian of the geodetic interval by 
\begin{equation}
    \nabla _{\mu }\nabla ^{\mu }\sigma = d - 2 \Delta ^{-\frac{1}{2}} \Delta ^{\frac{1}{2}}_{;\mu }\sigma ^{;\mu } ,
\end{equation}
which simplifies the recurrence relations.

The recurrence relations that we derive below for the Hadamard coefficients $U_{n}^{(d)}(x,x')$ and $V_{n}^{(d)}(x,x')$ can, at least in principle, be solved by integrating along the unique geodesic connecting the space-time points $x$ and $x'$.
As a result, these Hadamard coefficients are unique and determined by the space-time geometry, the background electromagnetic potential $A_{\mu }$ and the parameters $q$, $\xi $ and $m$ appearing in the charged scalar field equation (\ref{eq:inhom}). 
In particular, they do not depend on the state of the quantum scalar field, a point which is crucial for the process of Hadamard renormalization (see section~\ref{sec:expvals}).
In contrast,  the Hadamard coefficients $W_{n}^{(d)}(x,x')$ are not uniquely specified and the Hadamard coefficient $W_{0}^{(d)}(x,x')$ is completely undetermined by the recurrence relations.
If this coefficient were known, then the recurrence relations could (in theory) be solved for the coefficients $W_{n}^{(d)}(x,x')$ with $n\ge 1$ by integrating along the unique geodesic connecting $x$ and $x'$.
The undetermined Hadamard coefficient $W_{0}^{(d)}(x,x')$ depends on the state of the quantum scalar field as well as the space-time geometry and the electromagnetic potential (which we are regarding as fixed and purely classical).
 
Solving the recurrence relations by integrating along a geodesic is possible in practice only for space-times possessing a high degree of symmetry. 
For a general space-time and background electromagnetic potential, closed form expressions for the purely geometric Hadamard coefficients $U_{n}^{(d)}(x,x')$, $V_{n}^{(d)}(x,x')$  cannot be derived.
When performing renormalization (as we shall discuss in section~\ref{sec:expvals}), it is therefore useful to have covariant Taylor series expansions of the Hadamard coefficients $U_{n}^{(d)}(x,x')$, $V_{n}^{(d)}(x,x')$  (where applicable) in terms of $\sigma ^{;\mu }$. 
We write these covariant Taylor series expansions as follows:
\begin{subequations}
\label{eq:covTaylor}
\begin{eqnarray}
 U_{n}^{(d)}(x,x') & = & \sum _{j=0}^{\infty }
 U_{nj\alpha _{1}\ldots \alpha _{j}}^{(d)}(x)\sigma ^{;\alpha_{1}}(x,x')\ldots \sigma ^{;\alpha _{j}}(x,x'),
\label{eq:UcovTaylor}
 \\
 V_{n}^{(d)}(x,x') & = & \sum _{j=0}^{\infty }
 V_{nj\alpha _{1}\ldots \alpha _{j}}^{(d)}(x)\sigma ^{;\alpha_{1}}(x,x')\ldots \sigma ^{;\alpha _{j}}(x,x'),
 \label{eq:VcovTaylor}
\end{eqnarray}
\end{subequations}
where the $U_{nj}^{(d)}$, $V_{nj}^{(d)}$ are symmetric tensors of type $(0,j)$ defined at the space-time point $x$.
To find the $U_{nj}^{(d)}$, $V_{nj}^{(d)}$, we will require the covariant Taylor series expansions of the quantities $\sigma _{;\mu \nu }$, the square root of the Van Vleck-Morette determinant $\Delta  ^{\frac{1}{2}}$ and $\Delta ^{-\frac{1}{2}} \Delta ^{\frac{1}{2}}_{;\mu } \sigma ^{;\mu }$, which can be found in \cite{Decanini:2005gt}. To the order we require, the expansion for 
$\sigma _{;\mu \nu }$ is
\begin{eqnarray}
 \sigma _{;\mu \nu } & = & 
 g_{\mu \nu } - \frac{1}{3} R_{\mu \alpha _{1} \nu \alpha _{2}} \sigma ^{;\alpha _{1}} \sigma ^{;\alpha _{2}} 
 + \frac{1}{12} R_{\mu \alpha _{1}\nu \alpha _{2} ; \alpha _{3}} 
 \sigma^{;\alpha _{1}} \sigma ^{;\alpha _{2}} \sigma ^{;\alpha _{3}}
 \nonumber \\ & & 
 - \left[ 
 \frac{1}{60} R_{\mu \alpha _{1}\nu \alpha _{2} ; \alpha _{3}\alpha _{4}} +
 \frac{1}{45} R_{\mu \alpha _{1} \rho \alpha _{2}} R^{\rho }{}_{\alpha _{3} \nu \alpha _{4}} 
 \right] \sigma ^{;\alpha _{1}} \sigma ^{;\alpha _{2}} \sigma ^{;\alpha _{3}} \sigma ^{;\alpha _{4}}
 \nonumber \\ & & 
 + \ldots ,
\end{eqnarray}
while that for $\Delta ^{\frac{1}{2}}$ is
\begin{eqnarray}
 \Delta ^{\frac{1}{2}} & = & 
 1 + \frac{1}{12}R_{\alpha _{1}\alpha _{2}} \sigma ^{;\alpha _{1}} \sigma ^{;\alpha _{2}} 
 - \frac{1}{24} R_{\alpha _{1}\alpha _{2};\alpha _{3}} \sigma ^{;\alpha _{1}}\sigma ^{;\alpha _{2}} \sigma ^{;\alpha _{3}}
 \nonumber \\ & & 
 + \left[
 \frac{1}{80} R_{\alpha _{1}\alpha _{2};\alpha _{3}\alpha _{4}} 
 + \frac{1}{360} R^{\rho }{}_{\alpha _{1}\tau \alpha _{2}} R^{\tau }{}_{\alpha _{3}\rho \alpha _{4}} 
 \right. \nonumber \\ & & \left. 
 + \frac{1}{288} R_{\alpha _{1}\alpha _{2}} R_{\alpha _{3}\alpha _{4}} 
 \right] \sigma ^{;\alpha _{1}} \sigma ^{;\alpha _{2}} \sigma ^{;\alpha _{3}} \sigma ^{;\alpha _{4}}
 + \ldots ,
 \label{eq:Deltahalf}
\end{eqnarray}
and finally for $\Delta ^{-\frac{1}{2}} \Delta ^{\frac{1}{2}}_{;\mu } \sigma ^{;\mu }$
we have
\begin{eqnarray}
 \Delta ^{-\frac{1}{2}} \Delta ^{\frac{1}{2}}_{;\mu } \sigma ^{;\mu }
 & = & 
 \frac{1}{6} R_{\alpha _{1}\alpha _{2}} \sigma ^{;\alpha _{1}} \sigma ^{;\alpha _{2}} 
 - \frac{1}{24}R_{\alpha _{1}\alpha _{2};\alpha _{3}} \sigma ^{;\alpha _{1}}\sigma ^{;\alpha _{2}} \sigma ^{;\alpha _{3}}
 \nonumber \\ & & 
 + \left[  \frac{1}{120} R_{\alpha _{1}\alpha _{2};\alpha _{3}\alpha _{4}} + \frac{1}{90} R^{\rho }{}_{\alpha _{1}\tau \alpha _{2}} R^{\tau }{}_{\alpha _{3}\rho \alpha _{4}} 
 \right]  \sigma ^{;\alpha _{1}} \sigma ^{;\alpha _{2}} \sigma ^{;\alpha _{3}} \sigma ^{;\alpha _{4}}
 \nonumber \\ & &  + \ldots .
\end{eqnarray}
In deriving the covariant Taylor series expansions of the Hadamard coefficients, we will not assume the field equations for either the space-time curvature or the gauge field. 
The gauge field strength $F_{\mu \nu }$ is defined by 
\begin{equation}
    F_{\mu \nu } = \nabla _{\mu }A_{\nu } - \nabla _{\nu }A_{\mu }.
\end{equation}
The calculations below make use of the identity
\begin{equation}
    \nabla ^{\mu }\nabla ^{\nu} F_{\mu \nu } =0,
\end{equation}
which follows from the definition of the Ricci tensor and the antisymmetry of the gauge field strength $F_{\mu \nu }$.

\subsection{$d=2$}
\label{sec:d2}

\subsubsection{Recurrence relations for $d=2$}

To find the recurrence relations for the coefficients $V^{(2)}_{n}(x,x')$, we substitute the Hadamard parametrix (\ref{eq:GF_d2}) into the inhomogeneous scalar field equation (\ref{eq:inhom}). 
From the resulting terms proportional to $\log \left[ \sigma /\ell ^{2} + {\rm {i}}\epsilon \right]$, we find that the biscalar $V^{(2)}(x,x')$ satisfies the homogeneous scalar field equation (\ref{eq:scalar}):
\begin{equation}
0 = \left[ D_{\mu }D^{\mu } - \left( m^{2} + \xi R \right)  \right] V^{(2)},
\label{eq:V2eqn}
\end{equation}
while the remaining terms in the inhomogeneous scalar field equation (\ref{eq:inhom}) yield
\begin{equation}
    0   =   
    \sigma \left[ D_{\mu }D^{\mu } - \left( m^{2}+\xi R \right)  \right] W^{(2)} 
    + 2 \left[ \sigma ^{;\mu }D_{\mu } - \Delta ^{-\frac{1}{2}} \Delta ^{\frac{1}{2}}_{;\mu }\sigma ^{;\mu }  \right] V^{(2)}.
    \label{eq:W2eqn}
\end{equation}
We now substitute the power series expansions (\ref{eq:VWd2}) into (\ref{eq:V2eqn}, \ref{eq:W2eqn}) and compare coefficients. 
The ${\mathcal {O}}(\sigma ^{0})$ terms in (\ref{eq:W2eqn}) give
\begin{equation}
    0 =  \left[ \sigma ^{;\mu} D_{\mu } - \Delta ^{-\frac{1}{2}} \Delta ^{\frac{1}{2}}_{;\mu }\sigma ^{;\mu } \right] V^{(2)}_{0} .
    \label{eq:V20}
\end{equation}
In order that the leading-order singularity in the Hadamard parametrix (\ref{eq:GF_d2}) matches that in Minkowski space-time, it must be the case that $V^{(2)}_{0}(x,x')$ satisfies the boundary condition \cite{Decanini:2005eg}
\begin{equation}
    V^{(2)}_{0}(x,x)=-1.
    \label{eq:V20boundary}
\end{equation}
The recurrence relation satisfied by the coefficients $V^{(2)}_{n}$ is derived from (\ref{eq:V2eqn}) and takes the form
\begin{eqnarray}
    0 & = & 
    \left[ D_{\mu }D^{\mu } - \left( m^{2}+\xi R \right)  \right] V^{(2)}_{n}
    \nonumber \\  & & 
    +2\left( n+1 \right) \left[ \sigma ^{;\mu }D_{\mu }- \Delta ^{-\frac{1}{2}}\Delta ^{\frac{1}{2}}_{;\mu } \sigma ^{;\mu } + \left( 1 + n \right)   \right] V^{(2)}_{n+1},
    \label{eq:V2n}
\end{eqnarray}
for $n=0,1,\ldots $.
Finally, the higher-order terms in (\ref{eq:W2eqn}) give the following relation between the coefficients $W^{(2)}_{n}$ and $V^{(2)}_{n}$
\begin{eqnarray}
    0 & = & 
    \left[ D_{\mu } D^{\mu } - \left( m^{2}+\xi R \right)  \right] W^{(2)}_{n} 
    \nonumber \\ & & 
   + 2\left( n +1 \right) 
    \left[ \sigma ^{;\mu } D_{\mu } - \Delta ^{-\frac{1}{2}}\Delta ^{\frac{1}{2}}_{;\mu } \sigma ^{;\mu } + \left( 1 + n \right)  \right]  W^{(2)}_{n+1}
    \nonumber \\ & & 
    +2\left[ \sigma ^{;\mu } D_{\mu } - \Delta ^{-\frac{1}{2}}\Delta ^{\frac{1}{2}}_{;\mu } \sigma ^{;\mu } + 2\left( 1 + n \right) \right] V^{(2)}_{n+1}.
\end{eqnarray}

\subsubsection{Expansions for $d=2$}

The recurrence relations (\ref{eq:V2n}) together with the conditions (\ref{eq:V20}, \ref{eq:V20boundary}) uniquely determine the coefficients $V^{(2)}_{n}(x,x')$.
For a neutral scalar field, $V^{(2)}_{0}(x,x')$ is identically equal to $-\Delta ^{\frac{1}{2}}(x,x')$ \cite{Decanini:2005eg},  but the presence of the gauge potential $A_{\mu }$ in the covariant derivative in (\ref{eq:V20}) modifies $V^{(2)}_{0}(x,x')$ in the case where the scalar field is charged.
We therefore consider the covariant Taylor series expansions (\ref{eq:VcovTaylor}) of the coefficients $V^{(2)}_{n}(x,x')$. 
To find the RSET, it is sufficient to compute $V^{(2)}_{0}$ up to ${\mathcal {O}}(\sigma )$ and $V^{(2)}_{1}$ to zeroth order.

These expansions are known for a neutral scalar field \cite{Decanini:2005eg}, so we just need to find the corrections due to the gauge field potential $A_{\mu }$. 
Since the relations (\ref{eq:V20}, \ref{eq:V2n}) are linear in the coefficients $V^{(2)}_{n}$, we write
\begin{equation}
    V^{(2)}_{0} = -\Delta ^{\frac{1}{2}} + {\widetilde{V}}^{(2)}_{0}.
\end{equation}
Substituting in (\ref{eq:V20}), the equation satisfied by the correction ${\widetilde{V}}^{(2)}_{0}$ is
\begin{equation}
    0 = \left[ \sigma ^{;\mu} D_{\mu } - \Delta ^{-\frac{1}{2}} \Delta ^{\frac{1}{2}}_{;\mu }\sigma ^{;\mu } \right] {\widetilde {V}}^{(2)}_{0} + {\rm {i}}q A_{\mu }\sigma ^{;\mu }\Delta ^{\frac{1}{2}}.
    \label{eq:V20tilde}
\end{equation}
The boundary condition (\ref{eq:V20boundary}) implies that ${\widetilde {V}}^{(2)}_{00} =0$, and then from (\ref{eq:V20tilde}) we have
\begin{subequations}
\begin{eqnarray}
    {\widetilde {V}}^{(2)}_{01\mu } & = & -{\rm {i}}qA_{\mu },
    \\
    {\widetilde {V}}^{(2)}_{02\mu \nu } & = &\frac{{\rm {i}}q}{2} \nabla _{(\mu } A_{\nu )} +\frac{q^{2}}{2} A_{\mu }A_{\nu } = \frac{{\rm {i}}q}{2} D_{(\mu }A_{\nu )}.
\end{eqnarray}
\end{subequations}
Combining these with the expansion (\ref{eq:Deltahalf}) for $\Delta ^{\frac{1}{2}}$ and using the fact that, in two dimensions, the Ricci tensor and scalar are related by 
\begin{equation}
    R_{\mu \nu }= \frac{R}{2} g_{\mu \nu },
\end{equation}
we have
\begin{subequations}
\label{eq:V20expansion}
\begin{eqnarray}
    V^{(2)}_{00} & = & -1,
    \label{eq:V00d2}
     \\
    V^{(2)}_{01\mu } & = & -{\rm {i}}qA_{\mu },
    \label{eq:V01d2}
     \\ 
    V^{(2)}_{02\mu \nu } & = & - \frac{1}{24} R g_{\mu \nu } + \frac {{\rm {i}}q}{2} D_{(\mu }A_{\nu )}.
    \label{eq:V02d2}
\end{eqnarray}
\end{subequations}
We see that corrections to $V^{(2)}_{0}$ due to the electromagnetic potential arise at ${\mathcal {O}}(\sigma ^{\frac{1}{2}})$. 
The leading order corrections due to $A_{\mu }$ are independent of the space-time metric and therefore arise in flat Minkowski space-time as well as curved space-time.
At ${\mathcal {O}}(\sigma )$ in $V^{(2)}_{0}$, the corrections due to the electromagnetic potential do now depend on the metric via the covariant derivative.
Bearing in mind that the electromagnetic potential $A_{\mu }$ is real, it is straightforward to check that $\Re [V^{(2)}_{01\mu }]$ satisfies the condition (\ref{eq:rek1}) and $\Im [V^{(2)}_{02\mu \nu }]$ satisfies (\ref{eq:imk2}).

The next coefficient, $V^{(2)}_{1}$,  is found from the recurrence relation (\ref{eq:V2n}) with $n=0$.  
Since the form of this coefficient, ${\mathcal {V}}^{(2)}_{1}$, is known for a neutral scalar field \cite{Decanini:2005eg}, we write
\begin{equation}
    V^{(2)}_{1} = {\mathcal {V}}^{(2)}_{1} + {\widetilde {V}}^{(2)}_{1},
\end{equation}
where the correction ${\widetilde {V}}^{(2)}_{1}$ due to the gauge potential satisfies the equation
\begin{eqnarray}
0 & = & \left[ D_{\mu }D^{\mu } - \left( m^{2}+\xi R \right) \right] {\widetilde {V}}^{(2)}_{0} 
+ 2\left[ \sigma ^{;\mu }D_{\mu } - \Delta ^{-\frac{1}{2}}\Delta ^{\frac{1}{2}}_{;\mu }\sigma ^{;\mu } +1 \right] {\widetilde {V}}^{(2)}_{1}
\nonumber \\ & & 
-2{\rm {i}}q\sigma ^{;\mu }A_{\mu }{\mathcal {V}}^{(2)}_{1}
+ \left[ 2{\rm {i}}qA^{\mu }\nabla _{\mu }+{\rm {i}}q \left( \nabla _{\mu }A^{\mu }\right) + q^{2}A_{\mu }A^{\mu} \right] \Delta ^{\frac{1}{2}}.
\end{eqnarray}
To find the RSET, we only require the zeroth order term in ${\widetilde{V}}^{(2)}_{1}$, which is found to vanish
\begin{equation}
    {\widetilde{V}}^{(2)}_{10} = 0,
\end{equation}
and thus 
\begin{equation}
    V^{(2)}_{10} = {\mathcal {V}}^{(2)}_{10} = -\frac{1}{2} \left[ m^{2} + \left( \xi - \frac{1}{6} \right) R \right] .
    \label{eq:V1d2}
\end{equation}
Therefore, to leading order, the electromagnetic potential $A_{\mu }$
has no effect on $V^{(2)}_{1}$.

\subsection{$d>2$ even}
\label{sec:deven}

\subsubsection{Recurrence relations for general $d=2p$, $p>1$}

Substituting the Hadamard expansion (\ref{eq:GF_deven}, \ref{eq:UVWdeven}) into the inhomogeneous scalar field equation (\ref{eq:inhom}) and comparing coefficients of the various powers of $\sigma $, we find recurrence relations for the coefficients $U^{(2p)}_{n}(x,x')$,  $V^{(2p)}_{n}(x,x')$ and $W^{(2p)}_{n}(x,x')$.

The equation for $U^{(2p)}_{0}$ is derived from the ${\mathcal {O}}\left( \left[\sigma + {\rm {i}}\epsilon \right] ^{-p} \right) $ term in the inhomogeneous scalar field equation (\ref{eq:inhom}), and takes the form
\begin{equation}
     0 =  \left[ \Delta ^{-\frac{1}{2}} \Delta ^{\frac{1}{2}}_{;\mu } \sigma ^{;\mu }  - \sigma ^{;\mu }D _{\mu }  \right] U^{(2p)}_{0} .
     \label{eq:U2p0eqn}
\end{equation}
In order that the leading-order singularities in the Hadamard parametrix (\ref{eq:GF_deven}) match those in Minkowski space-time, it must be the case that $U^{(2p)}_{0}$ satisfies the boundary condition \cite{Decanini:2005eg}
\begin{equation}
    U^{(2p)}_{0} (x,x)=1.
    \label{eq:U2p0boundary}
\end{equation}
Due to the terms containing the electromagnetic potential $A_{\mu }$, it is not the case that $U_{0}^{(2p)}$ is equal to $\Delta ^{\frac{1}{2}}$ as for a neutral scalar field. Since the equation for $U_{0}^{(2p)}$ is linear, we can write
\begin{equation}
    U_{0}^{(2p)} = \Delta ^{\frac{1}{2}} + {\widetilde {U}}_{0}^{(2p)},
\end{equation}
and then ${\widetilde {U}}_{0}^{(2p)}$ satisfies the equation
\begin{equation}
\left[    \sigma ^{;\mu }D_{\mu } - \Delta ^{-\frac{1}{2}} \Delta ^{\frac{1}{2}}_{;\mu } \sigma ^{;\mu } \right] {\widetilde {U}}^{(2p)}_{0}  = {\rm {i}}qA_{\mu }\sigma ^{;\mu } \Delta ^{\frac{1}{2}} .
\label{eq:U0tilde}
\end{equation}
Terms in the inhomogeneous scalar field equation of order $\left[\sigma +{\rm {i}}\epsilon \right] ^{-p+n+1}$ for $n=0,1,\ldots p-3$ give the following recurrence relations for the Hadamard coefficients $U^{(2p)}_{n}(x,x')$ 
\begin{eqnarray}
   0 & = & 
   \left[ D_{\mu }D^{\mu } - \left( m^{2}+\xi R \right) \right] U^{(2p)}_{n}
   \nonumber \\ & & - 2\left( n+2-p\right) \left[ \Delta ^{-\frac{1}{2}} \Delta ^{\frac{1}{2}}_{;\mu } \sigma ^{;\mu }  - \sigma ^{;\mu }D _{\mu } -\left(n+1\right)   \right] U^{(2p)}_{n+1} ,
\end{eqnarray}
for $n=0,\ldots , p-3$.

The biscalar $V^{(2p)}(x,x')$ satisfies the homogeneous scalar field equation, derived from the terms in the inhomogeneous scalar field equation proportional to $\log \left[ \sigma /\ell ^{2}+{\rm {i}}\epsilon  \right] $:
\begin{equation}
 0 =     \left[ D_{\mu }D^{\mu } - \left( m^{2}+ \xi R \right) \right] V^{(2p)}(x,x') 
 \label{eq:V2eqndeven}
\end{equation}
and its expansion coefficients are governed by the recurrence relations
\begin{eqnarray}
   0 & = & 
   \left[  D_{\mu }D^{\mu } - \left( m^{2}+ \xi R \right)  \right] V^{(2p)}_{n} 
   \nonumber \\ & & 
   + 2 \left( n+1\right) \left[
   \sigma ^{;\mu }D _{\mu }- \Delta ^{-\frac{1}{2}} \Delta ^{\frac{1}{2}}_{;\mu } \sigma ^{;\mu }  + \left( p+n \right) 
    \right] V^{(2p)}_{n+1},
\end{eqnarray}
for $n=1,2,\ldots $.
The term of order $\left[\sigma + {\rm {i}}\epsilon \right] ^{-1}$ in the inhomogeneous scalar field equation (\ref{eq:inhom}) gives the boundary condition for $V^{(2p)}_{0}$:
\begin{eqnarray}
 0 & = &  \left[ D_{\mu }D^{\mu } - \left( m^{2}+\xi R\right)  \right] U^{(2p)}_{p-2} 
 \nonumber \\ & & 
 + 2 \left[ 
 \sigma ^{;\mu }D _{\mu }-\Delta ^{-\frac{1}{2}} \Delta ^{\frac{1}{2}}_{;\mu }\sigma ^{;\mu } +(p-1) 
 \right] V^{(2p)}_{0} . 
 \label{eq:V02p}
\end{eqnarray}

By considering the terms in the inhomogeneous scalar field equation (\ref{eq:inhom}) which are ${\mathcal {O}}\left(\left[ \sigma + {\rm {i}}\epsilon \right] ^{k} \right) $ for $k=-1,0,1,2,\ldots $, we find that the biscalars $V^{(2p)}(x,x')$ and $W^{(2p)}(x,x')$ are related via
\begin{eqnarray}
 0 & = & \left[ D_{\mu }D^{\mu } - \left( m^{2}+\xi R\right)  \right] U^{(2p)}_{p-2}
 + \sigma \left[ D_{\mu }D^{\mu } - \left( m^{2}+\xi R\right)  \right] W^{(2p)}
\nonumber \\ & & 
 + 2 \left[ 
 \sigma ^{;\mu }D _{\mu }-\Delta ^{-\frac{1}{2}} \Delta ^{\frac{1}{2}}_{;\mu }\sigma ^{;\mu } +(p-1) 
 \right] V^{(2p)}.
 \label{eq:VW2p}
\end{eqnarray}
The ${\mathcal {O}}\left(\sigma  ^{0}\right) $ contribution to (\ref{eq:VW2p}) is simply equation (\ref{eq:V02p}) for the Hadamard coefficient $V_{0}^{(2p)}(x,x')$.
The recurrence relations for the coefficients $W_{n}^{(2p)}(x,x')$ are found from the ${\mathcal {O}}\left( \left[ \sigma + {\rm {i}}\epsilon  \right] ^{n+1}\right) $ contributions to (\ref{eq:VW2p}) and are given by 
\begin{eqnarray}
 0 & = & 
 \left[ D_{\mu }D^{\mu } - \left( m^{2}+\xi R \right)  \right] W_{n}^{(2p)}
 \nonumber \\ & & 
 + 2\left( n+1 \right) \left[ 
 \sigma ^{;\mu }D _{\mu }- \Delta ^{-\frac{1}{2}} \Delta ^{\frac{1}{2}}_{;\mu }\sigma ^{;\mu }+ \left( n+p\right)  \right] W_{n+1}^{(2p)}
 \nonumber \\ & & 
 + 2 \left[ \sigma ^{;\mu }D_{\mu } - \Delta ^{-\frac{1}{2}} \Delta ^{\frac{1}{2}}_{;\mu }\sigma ^{;\mu } + \left( 2n + 1 + p \right)\right] V_{n+1}^{(2p)}.
\end{eqnarray}

\subsubsection{Expansions for $d=4$}

We begin by finding the covariant Taylor series expansion of ${\widetilde {U}}_{0}^{(4)}$ using the governing equation (\ref{eq:U0tilde}):
\begin{subequations}
\begin{eqnarray}
{\widetilde{U}}_{00}^{(4)}& = & 0;
 \\
 {\widetilde {U}}_{01\mu  }^{(4)} & = & {\rm {i}}qA_{\mu };
 \\
 {\widetilde {U}}_{02\mu \nu }^{(4)} & = & -\frac{1}{2}{\rm {i}}q \nabla _{(\mu } A_{\nu )}- \frac{1}{2}q^{2} A_{\mu} A_{\nu};
 \\
 {\widetilde {U}}_{03\mu \nu \lambda }^{(4)} & = & 
 \frac{1}{6} {\rm {i}}q \nabla _{(\mu }\nabla _{\nu } A_{\lambda )}
 + \frac{1}{12}{\rm{i}}q R_{(\mu \nu }A_{\lambda )} 
 + \frac{1}{2}q^{2} A_{(\mu }\nabla _{\nu }A_{\lambda )}
 \nonumber \\ & & 
 - \frac{1}{6} {\rm {i}}q^{3} A_{\mu } A_{\nu } A_{\lambda };
  \\
 {\widetilde {U}}_{04\mu \nu \lambda \tau }^{(4)} & = & 
 -\frac{1}{24}{\rm {i}}q A_{(\mu }\nabla _{\nu }R_{\lambda  \tau )}
 - \frac{1}{24}{\rm {i}}q R_{(\mu \nu }\nabla _{\lambda }A_{\tau )}
 - \frac{1}{24}{\rm {i}}q \nabla _{(\mu }\nabla _{\nu }\nabla _{\lambda } A_{\tau )}
 \nonumber \\ & & 
 - \frac{1}{8}q^{2} \left( \nabla _{(\mu } A_{\nu } \right) \left( \nabla _{\lambda }A_{\tau )} \right)
 - \frac{1}{24}q^{2} R_{(\mu \nu }A_{\lambda }A_{\tau )}
 - \frac{1}{6}q^{2} A_{(\mu }\nabla _{\nu }\nabla _{\lambda }A_{\tau )}
 \nonumber \\ & & 
 + \frac{1}{4}{\rm {i}}q^{3} A_{(\mu }A_{\nu }\nabla _{\lambda }A_{\tau )}
 + \frac{1}{24}q^{4} A_{\mu }A_{\nu }A_{\lambda }A_{\tau }.
\end{eqnarray}
\end{subequations}
Combining these with the expansion of $\Delta ^{\frac{1}{2}}$ (\ref{eq:Deltahalf}), and simplifying, we find compact expressions in terms of the covariant derivative:
\begin{subequations}
\label{eq:Ud4}
\begin{eqnarray}
 U^{(4)}_{00} & = & 1; 
 \\
 U^{(4)}_{01\mu } & = & {\rm {i}}qA_{\mu };
 \label{eq:U401}
 \\
 U^{(4)}_{02\mu \nu } & = & 
 \frac{1}{12} R_{\mu \nu } - \frac{{\rm {i}}q}{2} D_{(\mu }A_{\nu )};
 \\
 U^{(4)}_{03\mu \nu \lambda } & = & 
 -\frac{1}{24} R_{( \mu \nu ; \lambda )}
 + \frac{{\rm {i}}q}{6}D_{(\mu }D_{\nu }A_{\lambda )}
+ \frac{{\rm {i}}q}{12} A_{(\mu }R_{\nu  \lambda )};
\\
 U^{(4)}_{04\mu \nu \lambda \tau } & = & 
 \frac{1}{80} R_{(\mu \nu ; \lambda \tau )} + \frac{1}{360}R^{\rho }{}_{(\mu | \psi | \nu }R^{\psi }{}_{\lambda | \rho | \tau )} 
 + \frac{1}{288} R_{(\mu \nu }R_{\lambda \tau )}
 \nonumber \\ & & 
-\frac{{\rm {i}}q}{24} D_{(\mu }D_{\nu }D_{\lambda }A_{\tau )} - \frac{{\rm {i}}q}{24}D_{(\mu }\left[ A_{\nu }R_{\lambda \tau )} \right] .
\end{eqnarray}
\end{subequations}
As in $d=2$, we find corrections to $U^{(4)}_{0}$ due to the electromagnetic potential at ${\mathcal {O}}(\sigma ^{\frac{1}{2}})$, and the leading-order corrections due to $A_{\mu }$ are independent of the space-time metric.
Terms involving coupling between the electromagnetic potential and the Ricci curvature arise at ${\mathcal {O}}(\sigma ^{\frac{3}{2}})$. 
It is straightforward to check that $\Im [U^{(4)}_{02\mu \nu }]$ satisfies (\ref{eq:imk2}), while (\ref{eq:rek1k3}) also holds.

To find $V_{0}^{(4)}$, we proceed in a similar way and write
\begin{equation}
    V_{0}^{(4)}= {\mathcal {V}}_{0}^{(4)}+ {\widetilde{V}}_{0}^{(4)},
\end{equation}
where ${\mathcal {V}}_{0}^{(4)}$ is the form of $V_{0}^{(4)}$ for an uncharged scalar field, as given in \cite{Decanini:2005eg}. 
Then the equation satisfied by ${\widetilde {V}}_{0}^{(4)}$ is 
\begin{eqnarray}
& & \hspace{-0.5cm} 2\left[ \sigma ^{;\mu } \nabla _{\mu } - \Delta ^{-\frac{1}{2}} \Delta ^{\frac{1}{2}}_{;\mu } \sigma ^{;\mu } - {\rm {i}}q A_{\mu }\sigma ^{;\mu } +1  \right] {\widetilde {V}}_{0}^{(4)}
  + \left[ D_{\mu }D^{\mu } - \left( m^{2}+\xi R \right)  \right] {\widetilde {U}}_{0}^{(4)} 
  \nonumber \\  & & 
  = 2{\rm {i}}qA_{\mu }\sigma ^{;\mu }{\mathcal {V}}_{0}^{(4)} + 
  \left[ 2{\rm {i}}q A^{\mu} \nabla _{\mu } + {\rm {i}}q \left( \nabla _{\mu }A^{\mu } \right) + q^{2} A_{\mu }A^{\mu } \right] \Delta ^{\frac{1}{2}}.
\end{eqnarray}
As an intermediate step, we can write $D_{\mu }D^{\mu }{\widetilde{U}}_{0}^{(4)}$ in the form
\begin{eqnarray}
D_{\mu }D^{\mu }{\widetilde{U}}_{0}^{(4)} &  = & 
{\rm {i}}qD^{\mu }A_{\mu } 
+ \left[ 6g^{\alpha \beta} {\widetilde{U}}_{03\alpha \beta \mu }^{(4)}
-{\rm {i}}qD^{\alpha }D_{\mu }A_{\alpha }
+\frac{{\rm {i}}q}{3}A^{\alpha }R_{\alpha \mu } 
\right] \sigma ^{;\mu }
\nonumber \\ & & 
+ \left[ 
12 g^{\alpha \beta } {\widetilde {U}}^{(4)}_{04\alpha \beta \mu \nu }+ 6D^{\alpha }{\widetilde {U}}_{03\alpha \mu \nu }^{(4)} + D^{\alpha }D_{\alpha }{\widetilde {U}}^{(4)}_{02\mu \nu }
\right. \nonumber \\ & & \left.
+ \frac{2}{3} {\widetilde{U}}_{02\alpha ( \mu }^{(4)}R_{\nu )}{}^{\alpha }
- \frac{{\rm {i}}q}{12}A^{\alpha }R_{\alpha (\mu ; \nu )}- \frac{{\rm {i}}q}{4} A^{\beta }R_{\beta (\mu | \alpha | \nu ) }{}^{;\alpha }
\right] \sigma ^{;\mu }\sigma ^{;\nu }
\nonumber \\  & & +\ldots .
\end{eqnarray}
This simplifies to 
\begin{eqnarray}
D_{\mu }D^{\mu }{\widetilde{U}}_{0}^{(4)} &  = & 
{\rm {i}}qD^{\mu }A_{\mu } 
+ \frac{{\rm {i}}q}{3} \left[ A^{\alpha }R_{\alpha \mu } + \frac{1}{2}RA_{\mu }
+ \nabla ^{\alpha }F_{\alpha \mu }
\right] \sigma ^{;\mu }
\nonumber \\ & & 
+\left[ 
\frac{{\rm {i}}q}{12} R_{\mu \nu }D_{\alpha }A^{\alpha }
- \frac{{\rm {i}}q}{12} R D_{(\mu }A_{\nu )}
+ \frac{{\rm {i}}q}{12} A^{\alpha }R_{\mu \nu ;\alpha}
\right. \nonumber \\ & & \left.
- \frac{{\rm {i}}q}{6}A^{\alpha }R_{\alpha (\mu ;\nu )}
+ \frac{q^{2}}{4}F^{\alpha }{}_{\mu }F_{\nu \alpha }
+ \frac{{\rm {i}}q}{12}\nabla ^{\alpha }\nabla _{(\mu }F_{\nu )\alpha }
\right. \nonumber \\ & & \left.
+ \frac{q^{2}}{3}A_{(\mu }\nabla ^{\alpha }F_{\nu )\alpha }
- \frac{{\rm {i}}q}{12}R^{\alpha }{}_{(\mu }F_{\nu ) \alpha }
\right] \sigma ^{;\mu }\sigma ^{;\nu } + \ldots 
\end{eqnarray}
The covariant Taylor series expansion of ${\widetilde{V}}_{0}^{(4)}$ is then
\begin{subequations}
\begin{eqnarray}
{\widetilde{V}}_{00}^{(4)} & = & 0;
 \\
{\widetilde {V}}_{01\mu }^{(4)} & = & 
\frac{{\rm {i}}q}{2} \left[ m^{2}+ \left(  \xi - \frac{1}{6}\right)R \right] A_{\mu }- \frac{{\rm {i}}q}{12} \nabla ^{\alpha }F_{\alpha \mu };
\\
{\widetilde {V}}_{02\mu \nu }^{(4)} & = & 
-\frac{{\rm {i}}q}{4} \left[ m^{2}+ \left(  \xi - \frac{1}{6}\right)R \right] D_{(\mu }A_{\nu )} - \frac{{\rm {i}}q}{4} \left( \xi - \frac{1}{6} \right) A_{(\mu }R_{;\nu )}
\nonumber \\ & &
- \frac{q^{2}}{24}F^{\alpha }{}_{\mu }F_{\nu \alpha }
- \frac{q^{2}}{12} A_{(\mu  }\nabla ^{\alpha }F_{\nu ) \alpha }
- \frac{{\rm {i}}q}{24} \nabla _{(\mu }\nabla ^{\alpha } F_{\nu ) \alpha }.
\end{eqnarray}
\end{subequations}
Combining these with the expansion of ${\mathcal {V}}_{0}^{(4)}$ given in \cite{Decanini:2005eg}, we have
\begin{subequations}
\label{eq:V40pexpansion}
\begin{eqnarray}
V_{00}^{(4)} & = & 
\frac{1}{2} \left[ m^{2}+ \left( \xi - \frac{1}{6} \right) R \right] ;
\label{eq:V00d4}
\\
V_{01\mu }^{(4)} & = & 
-\frac{1}{4}\left( \xi - \frac{1}{6}\right) R_{;\mu } 
+ \frac{{\rm {i}}q}{2} \left[ m^{2}+ \left(  \xi - \frac{1}{6}\right)R \right] A_{\mu }
\nonumber \\ & & 
- \frac{{\rm {i}}q}{12} \nabla ^{\alpha }F_{\alpha \mu };
\label{eq:V01d4}
\\
V_{02\mu \nu }^{(4)} & = & 
\frac{1}{24} \left[ m^{2}+\left( \xi - \frac{1}{6} \right) R \right] R_{\mu \nu }
+ \frac{1}{12} \left( \xi - \frac{3}{20} \right) R_{;\mu \nu }
- \frac{1}{240} \Box R_{\mu \nu }
\nonumber \\  & & 
+ \frac{1}{180}R^{\alpha }{}_{\mu }R_{\alpha \nu }
- \frac{1}{360} R^{\alpha \beta }R_{\alpha \mu \beta \nu }
- \frac{1}{360} R^{\alpha \beta \gamma }{}_{\mu} R _{\alpha \beta \gamma \nu }
\nonumber \\ & & 
-\frac{{\rm {i}}q}{4} \left[ m^{2}+ \left(  \xi - \frac{1}{6}\right)R \right] D_{(\mu }A_{\nu )} - \frac{{\rm {i}}q}{4} \left( \xi - \frac{1}{6} \right) A_{(\mu }R_{;\nu )}
\nonumber \\ & & 
- \frac{q^{2}}{24}F^{\alpha }{}_{\mu }F_{\nu \alpha }
- \frac{q^{2}}{12} A_{(\mu  }\nabla ^{\alpha }F_{\nu ) \alpha }
- \frac{{\rm {i}}q}{24} \nabla _{(\mu }\nabla ^{\alpha } F_{\nu ) \alpha }.
\label{eq:V02d4}
\end{eqnarray}
\end{subequations}
The leading order form of $V^{(4)}_{0}$ is therefore, as in $d=2$, unaffected by the presence of the gauge potential $A_{\mu }$. 
At ${\mathcal {O}}(\sigma ^{\frac{1}{2}})$, as well as corrections due to the potential $A_{\mu }$, the electromagnetic field strength $F_{\mu \nu }$ also arises, with couplings between the gauge potential and curvature appearing at ${\mathcal {O}}(\sigma ^{1})$.
Again, the coefficients in the covariant Taylor series expansion of $V_{0}^{(4)}$ satisfy (\ref{eq:rek1}, \ref{eq:imk2}).

Similarly, we write $V_{1}^{(4)}$ as
\begin{equation}
        V_{1}^{(4)}= {\mathcal {V}}_{1}^{(4)}+ {\widetilde{V}}_{1}^{(4)},
\end{equation}
where ${\mathcal {V}}_{1}^{(4)}$ is the form of $V_{1}^{(4)}$ for an uncharged scalar field, as given in \cite{Decanini:2005eg}. 
The equation satisfied by ${\widetilde {V}}_{1}^{(4)}$ is
\begin{eqnarray}
& & \hspace{-0.5cm} 2\left[ \sigma ^{;\mu } \nabla _{\mu } - \Delta ^{-\frac{1}{2}} \Delta ^{\frac{1}{2}}_{;\mu } \sigma ^{;\mu } - {\rm {i}}q A_{\mu }\sigma ^{;\mu } +2  \right] {\widetilde {V}}_{1}^{(4)}
  + \left[ D_{\mu }D^{\mu } - \left( m^{2}+\xi R \right)  \right] {\widetilde {V}}_{0}^{(4)} 
  \nonumber \\  & & 
  = 2{\rm {i}}qA_{\mu }\sigma ^{;\mu }{\mathcal {V}}_{1}^{(4)} + 
  \left[ 2{\rm {i}}q A^{\mu} \nabla _{\mu } + {\rm {i}}q \left( \nabla _{\mu }A^{\mu } \right) + q^{2} A_{\mu }A^{\mu } \right] {\mathcal {V}}_{0}^{(4)}.
\end{eqnarray}
Only the zeroth order term in ${\widetilde {V}}_{1}^{(4)}$ is needed for a computation of the RSET in $d=4$.
We find that this is given by 
\begin{equation}
    {\widetilde {V}}_{10}^{(4)} = -\frac{q^{2}}{48}F^{\alpha \beta }F_{\alpha \beta }.
\end{equation}
Combining this with the expression for ${\mathcal {V}}_{1}^{(4)}$ from \cite{Decanini:2005eg}, we have
\begin{eqnarray}
    V^{(4)}_{10} & = & \frac{1}{8} \left[ m^{2} +\left( \xi - \frac{1}{6} \right) R \right] ^{2} - \frac{1}{24} \left( \xi - \frac{1}{5} \right) \Box R - \frac{1}{720} R^{\alpha \beta }R_{\alpha \beta }
    \nonumber \\ & & 
    + \frac{1}{720} R^{\alpha \beta \gamma \delta}R_{\alpha \beta \gamma \delta } 
    - \frac{q^{2}}{48} F^{\alpha \beta }F_{\alpha \beta }. 
    \label{eq:V1d4}
\end{eqnarray}
Here there is a correction due to the electromagnetic field strength at leading order. 
This correction will be important in Section \ref{sec:trace} where we consider the trace anomaly.

\subsection{$d$ odd}
\label{sec:dodd}

\subsubsection{Recurrence relations for general $d=2p+1$}

The recurrence relations satisfied by the coefficients in the power series expansions (\ref{eq:UWdodd}) for odd numbers of space-time dimensions are derived in a similar way to those for even space-time dimensions.
First we substitute the Hadamard parametrix (\ref{eq:GF_dodd}) into the inhomogeneous scalar field equation (\ref{eq:inhom}). 
Due to the fractional power of $\sigma $ in (\ref{eq:GF_dodd}), the resulting equation consists of two parts.
The first involves only integer powers of $\sigma $ and shows that $W^{(2p+1)}$ satisfies the homogeneous scalar field equation
\begin{equation}
    0 = \left[ D_{\mu }D^{\mu } - \left( m^{2}+\xi R \right)  \right] W^{(2p+1)},
    \label{eq:W2p+1eqn}
\end{equation}
while the second involves noninteger powers of $\sigma $ and gives the equation satisfied by $U^{(2p+1)}$, namely
\begin{eqnarray}
 0 & = &     
 \sigma \left[ D_{\mu }D^{\mu } - \left( m^{2}+\xi R \right) \right] U^{(2p+1)}
 \nonumber \\ & & 
 - \left( 2p-1 \right)  \left[ 
 \sigma ^{;\mu }D_{\mu } - \Delta ^{-\frac{1}{2}}\Delta ^{\frac{1}{2}}_{;\mu } \sigma ^{;\mu }
 \right] U^{(2p+1)} .
 \label{eq:U2p+1eqn}
\end{eqnarray}
Substituting the power series expansion (\ref{eq:UWdodd}) for $U^{(2p+1)}$ into (\ref{eq:U2p+1eqn}), the ${\mathcal {O}}(\sigma ^{0})$ term gives the equation satisfied by $U^{(2p+1)}_{0}$:
\begin{equation}
0 = \left[ \sigma ^{;\mu }D_{\mu } - \Delta ^{-\frac{1}{2}}\Delta ^{\frac{1}{2}}_{;\mu } \sigma ^{;\mu } \right] U^{(2p+1)}_{0},
\label{eq:U2p+10eqn}
\end{equation}
while the coefficient of $\sigma ^{n+1}$ yields the recurrence relation
\begin{eqnarray}
 0 & = & 
 \left[ D_{\mu }D^{\mu } - \left( m^{2} + \xi R \right) \right] U^{(2p+1)}_{n}
 \nonumber \\  & & 
 + \left( 2n+3 -2p \right) \left[ \sigma ^{;\mu }D_{\mu } - \Delta ^{-\frac{1}{2}} \Delta ^{\frac{1}{2}}_{;\mu } \sigma ^{;\mu } + \left( n+1 \right) \right] U^{(2p+1)}_{n+1}. 
\end{eqnarray}
The boundary condition satisfied by $U^{(2p+1)}_{0}$ is \cite{Decanini:2005eg}
\begin{equation}
    U^{(2p+1)}_{0}(x,x)=1.
    \label{eq:U2p+10boundary}
\end{equation}

\subsubsection{Expansions for $d=3$}

We first note that the equation (\ref{eq:U2p+10eqn}) and boundary condition (\ref{eq:U2p+10boundary}) satisfied by $U^{(2p+1)}_{0}$ are identical to those (\ref{eq:U2p0eqn}, \ref{eq:U2p0boundary}) satisfied by $U^{(2p)}_{0}$.
Therefore the expansion of $U^{(3)}_{0}$ is the same as that of $U^{(4)}_{0}$, and the terms required to calculate the RSET are:
\begin{subequations}
\begin{eqnarray}
     U^{(3)}_{00} & = & 1; 
 \\
 U^{(3)}_{01\mu } & = & {\rm {i}}qA_{\mu };
  \\
 U^{(3)}_{02\mu \nu } & = & 
 \frac{1}{12} R_{\mu \nu } - \frac{{\rm {i}}q}{2} D_{(\mu }A_{\nu )};
  \\
 U^{(3)}_{03\mu \nu \lambda } & = & 
 -\frac{1}{24} R_{( \mu \nu ; \lambda )}
 + \frac{{\rm {i}}q}{6}D_{(\mu }D_{\nu }A_{\lambda )}
+ \frac{{\rm {i}}q}{12} A_{(\mu }R_{\nu  \lambda )}.
\end{eqnarray}
\end{subequations}
In $d=3$, we also require $U^{(3)}_{1}$ to ${\mathcal {O}}(\sigma ^{;\mu })$ in order to find the RSET.
The equation satisfied by $U^{(3)}_{1}$ is
\begin{equation}
  0  =
 \left[ D_{\mu }D^{\mu } - \left( m^{2} + \xi R \right) \right] U^{(3)}_{0}
 +  \left[ \sigma ^{;\mu }D_{\mu } - \Delta ^{-\frac{1}{2}} \Delta ^{\frac{1}{2}}_{;\mu } \sigma ^{;\mu } + 1 \right] U^{(3)}_{1}.     
\end{equation}
This is very similar to the equation (\ref{eq:V02p}) for $V^{(4)}_{0}$ in four dimensions:
\begin{equation}
  0  =  \left[ D_{\mu }D^{\mu } - \left( m^{2}+\xi R\right)  \right] U^{(4)}_{0} 
 + 2 \left[ 
 \sigma ^{;\mu }D _{\mu }-\Delta ^{-\frac{1}{2}} \Delta ^{\frac{1}{2}}_{;\mu }\sigma ^{;\mu } +1 
 \right] V^{(4)}_{0} .   
\end{equation}
We can therefore easily deduce the expansion of $U^{(3)}_{1}$ from that 
(\ref{eq:V40pexpansion}) of $V^{(4)}_{0}$. 
The terms required for the computation of the RSET are therefore
\begin{subequations}
\begin{eqnarray}
     U^{(3)}_{10} & = & m^{2}+\left(  \xi - \frac{1}{6}\right)  R ,
      \\
     U^{(3)}_{11\mu } & = & 
     -\frac{1}{2}\left( \xi - \frac{1}{6}\right) R_{;\mu } 
+ {\rm {i}}q \left[ m^{2}+ \left(  \xi - \frac{1}{6}\right)R \right] A_{\mu }- \frac{{\rm {i}}q}{6} \nabla ^{\alpha }F_{\alpha \mu }.
\end{eqnarray}
\end{subequations}

\section{Renormalized expectation values}
\label{sec:expvals}

In (\ref{eq:GF_d2}, \ref{eq:GF_deven}, \ref{eq:GF_dodd}) we have written down the Hadamard parametrix for the Feynman Green's function $G_{\rm {F}}(x,x')$, which depends on three sesquisymmetric biscalars $U^{(d)}(x,x')$, $V^{(d)}(x,x')$ and $W^{(d)}(x,x')$. 
For a neutral scalar field, the biscalars $U^{(d)}(x,x')$ and $V^{(d)}(x,x')$ are purely geometric and are uniquely determined by the space-time geometry. 
Here we are considering a charged scalar field interacting not only with the space-time geometry but also with a fixed, background, purely classical electromagnetic field. 
As a result, the biscalars $U^{(d)}(x,x')$ and $V^{(d)}(x,x')$ depend on the electromagnetic potential as well as the space-time metric, but they are still uniquely determined by the recurrence relations derived in the previous section.

The part of the Feynman Green's function which is divergent in the coincidence limit is $G_{\rm {S}}(x,x')$, where we define
\begin{equation}
    -{\rm {i}}G_{\rm {S}}(x,x') = \begin{cases}
    {\displaystyle {\alpha ^{(2)} V^{(2)}(x,x') \ln \left[ \frac{\sigma (x,x')}{\ell ^{2}} + {\rm {i}} \epsilon \right]  }} &  d=2,
    \\
    {\displaystyle {\alpha ^{(2p)} \left\{ 
    \frac{U^{(2p)}(x,x')}{\left[ \sigma (x,x') +{\rm {i}} \epsilon \right]  ^{p-1}} 
    + V^{(2p)}(x,x') \ln \left[ \frac{\sigma (x,x')}{\ell ^{2}} + {\rm {i}} \epsilon \right] 
    \right\} }}  & d=2p,
    \\
    {\displaystyle {\alpha ^{(2p+1)} \frac{U^{(2p+1)}(x,x')}{\left[ \sigma (x,x')+ {\rm {i}} \epsilon  \right] ^{p-\frac{1}{2}} } }} & d=2p+1.
    \end{cases}
\end{equation}
Since $G_{\rm {S}}(x,x')$ depends only on the biscalars $U^{(d)}(x,x')$ and $V^{(d)}(x,x')$, it is uniquely determined by the background space-time and electromagnetic potential, and takes the same form independent of the quantum state of the scalar field.
In this section we turn to the construction of renormalized expectation values of the scalar field condensate, current and stress-energy tensor.
We first subtract the divergent $G_{\rm {S}}(x,x')$ from the Feynman Green's function $G_{\rm {F}}(x,x')$ to give a regularized Green's function $G_{{\rm {R}}}(x,x')$, which takes the form
\begin{equation}
    -{\rm {i}}G_{\rm {R}}(x,x') =-{\rm {i}}\left[ G_{\rm {F}}(x,x') - G_{\rm {S}}(x,x') \right] = \alpha ^{(d)}W^{(d)}(x,x')
    \label{eq:Greg}
\end{equation}
for any number of space-time dimensions $d$.
The renormalized current and RSET will be given by applying appropriate differential operators to $-{\rm {i}}G_{{\rm {R}}}(x,x')$ and then taking the coincidence limit $x'\rightarrow x$.
In this section we first derive some useful identities for the biscalar $W^{(d)}(x,x')$, before turning our attention to the renormalized expectation values themselves.
We emphasize that the biscalar $W^{(d)}(x,x')$, and therefore the regularized Green's function $G_{\rm {R}}(x,x')$ are not uniquely determined by the background geometry and electromagnetic potential, and depend on the details of the quantum state of the scalar field.
This means that the expectation values we study in this section will also depend on the quantum state of the field.
Our discussion in this section is for a general space-time and background electromagnetic potential. 
We therefore do not specify a quantum state for the field, and as a consequence the biscalar $W^{(d)}(x,x')$ will remain undetermined in our analysis.
We focus on general properties of the renormalized expectation values arising from Hadamard renormalization.
In practical applications, a particular space-time, electromagnetic potential and quantum state will be specified, which will enable $W^{(d)}(x,x')$, and therefore renormalized expectation values, to be computed explicitly using the results presented in this section.

\subsection{Properties of the biscalar $W^{(d)}(x,x')$}
\label{sec:W}

The biscalar $W^{(d)}(x,x')$ appearing in the Hadamard parametrix (\ref{eq:GF_d2}, \ref{eq:GF_deven}, \ref{eq:GF_dodd}) cannot be uniquely determined from recurrence relations derived in section~\ref{sec:Hadamard}, and depends on the quantum state under consideration as well as the background geometry and electromagnetic potential. 
We therefore leave this biscalar undetermined, but are able nonetheless to derive some of its properties.
Following \cite{Decanini:2005eg}, we write $W^{(d)}(x,x')$ as a covariant Taylor series expansion in the form
\begin{equation}
    W^{(d)}(x,x') = w^{(d)}_{0}(x) + w^{(d)}_{1\mu }(x) \sigma ^{;\mu }+ w^{(d)}_{2\mu \nu }(x)\sigma ^{;\mu }\sigma ^{;\nu } + 
    w^{(d)}_{3\mu \nu \lambda }(x)\sigma ^{;\mu }\sigma ^{;\nu }\sigma ^{;\lambda }+ \ldots ,
    \label{eq:WTaylor}
\end{equation}
where the coefficients $w^{(d)}_{0}$, $w^{(d)}_{1\mu }$, $w^{(d)}_{2\mu \nu }$ and $w^{(d)}_{3\mu \nu \lambda }$ depend only on the space-time point $x$.
We emphasize that, since the biscalar $W^{(d)}(x,x')$ depends on the quantum state of the field, so too do the coefficients $w^{(d)}_{0}$, $w^{(d)}_{1\mu }$, $w^{(d)}_{2\mu \nu }$ and $w^{(d)}_{3\mu \nu \lambda }$, and therefore they are undetermined by our general analysis. 
Since $W^{(d)}(x,x')$ is sesquisymmetric, the coefficients $w^{(d)}_{1\mu }$, $w^{(d)}_{2\mu \nu }$ and $w^{(d)}_{3\mu \nu \lambda }$ will satisfy the conditions (\ref{eq:rek1k3}, \ref{eq:imk2}).
Following \cite{Decanini:2005eg}, we now derive some identities involving the coefficients in (\ref{eq:WTaylor}) which are useful for the study of the general properties of the expectation values of the current and stress-energy tensor.

First consider the case of even numbers of space-time dimensions. 
The equations (\ref{eq:W2eqn}, \ref{eq:VW2p}) satisfied by $W(x,x')$ in this case take the form
\begin{equation}
    0 = 
    \left[ D_{\mu } D^{\mu } - \left( m^{2}+\xi R\right) \right] W^{(2p)}
    + 2\left[ \sigma ^{;\mu }D_{\mu }+  \left( p+1 \right)  \right] V^{(2p)}_{1} + {\mathcal {O}}(\sigma ).
\end{equation}
Inserting the expansions (\ref{eq:VcovTaylor}, \ref{eq:WTaylor}),  we find the following identities
\begin{subequations}
\label{eq:Widentities}
\begin{eqnarray}
0 & = & 
D_{\mu }D^{\mu }w^{(2p)}_{0} + 2 D^{\mu } w^{(2p)}_{1\mu} + 2g^{\mu \nu}w^{(2p)}_{2\mu \nu }- \left( m^{2}+\xi R \right) w^{(2p)}_{0} 
\nonumber \\ & & 
+ 2\left( p+1 \right) V^{(2p)}_{10},
\label{eq:Widentity1} \\
0 & = & 
D_{\mu }D^{\mu }w^{(2p)}_{1\alpha } + 4D^{\mu }w^{(2p)}_{2\alpha \mu } + 6 g^{\mu \nu }w^{(2p)}_{3\alpha \mu \nu } + \frac{1}{3}R^{\mu }{}_{\alpha } w^{(2p)}_{1\mu }
- \left( m^{2}+\xi R \right) w^{(2p)}_{1 \alpha }
\nonumber \\ & & 
+ 2D_{\alpha }V^{(2p)}_{10} + 2\left( p+2\right) V^{(2p)}_{11\alpha }.
\label{eq:Widentity2}
\end{eqnarray}
\end{subequations}
The real and imaginary parts of (\ref{eq:Widentity1}) are (having simplified using (\ref{eq:rek1}, \ref{eq:imk2}))
\begin{subequations}
\label{eq:evenw1}
\begin{eqnarray}
0 & = & 
2g^{\mu \nu }\Re \left( w^{(2p)}_{2\mu \nu }\right) +2qA^{\mu }\Im \left( w^{(2p)}_{1\mu } \right) -\left[ m^{2}+\xi R + q^{2} A_{\mu }A^{\mu }  \right]  w^{(2p)}_{0} 
\nonumber \\ & & + 2\left( p+1 \right) V^{(2p)}_{10},
\label{eq:evenw1re} \\
0 & = & \nabla ^{\mu} \Im \left( w^{(2p)}_{1\mu } \right) - qA^{\mu } \nabla _{\mu }w^{(2p)}_{0} - q \left( \nabla ^{\mu }A_{\mu } \right) w_{0}^{(2p)}.
\label{eq:evenw1im}
\end{eqnarray}
\end{subequations}
We only require the real part of (\ref{eq:Widentity2}).
Using (\ref{eq:rek1k3}, \ref{eq:imk2}), this simplifies to
\begin{eqnarray}
0 & = & 
2\nabla ^{\mu } \Re \left( w^{(2p)}_{2\alpha \mu } \right)
+qA^{\mu }\nabla _{\mu }\Im \left( w^{(2p)}_{1\alpha }\right)
+ 2q \left( \nabla _{(\alpha }A^{\mu } \right) \Im \left( w^{(2p)}_{1\mu )}\right)
\nonumber \\ & & 
- \frac{1}{4} \nabla _{\alpha } \Box w^{(2p)}_{0}
- \frac{1}{2} R^{\mu }{}_{\alpha }w^{(2p)}_{0;\mu }
- \left[ \frac{1}{2}\xi R_{;\alpha } + q^{2}A^{\mu }\nabla _{\alpha }A_{\mu } \right] w^{(2p)}_{0}
\nonumber \\ & & 
+ \nabla _{\alpha }V^{(2p)}_{10}.
\label{eq:evenw2}
\end{eqnarray}

The calculation proceeds similarly for odd numbers of space-time dimensions. In this case $W^{(2p+1)}$ satisfies the homogeneous scalar field equation (\ref{eq:W2p+1eqn}) and the identities (\ref{eq:Widentities}) simplify:
\begin{subequations}
\begin{eqnarray}
0 & = & 
D_{\mu }D^{\mu }w^{(2p+1)}_{0} + 2 D^{\mu } w^{(2p+1)}_{1\mu} + 2g^{\mu \nu}w^{(2p+1)}_{2\mu \nu }- \left( m^{2}+\xi R \right) w^{(2p+1)}_{0} ,
 \\
0 & = & 
D_{\mu }D^{\mu }w^{(2p+1)}_{1\alpha } + 4D^{\mu }w^{(2p+1)}_{2\alpha \mu } + 6 g^{\mu \nu }w^{(2p+1)}_{3\alpha \mu \nu } + \frac{1}{3}R^{\mu }{}_{\alpha } w^{(2p+1)}_{1\mu } 
\nonumber \\ & & 
- \left( m^{2}+\xi R \right) w^{(2p+1)}_{1 \alpha }.
\end{eqnarray}
\end{subequations}
The results corresponding to those in (\ref{eq:evenw1}, \ref{eq:evenw2})
are
\begin{subequations}
\label{eq:oddw1}
\begin{eqnarray}
0 & = & 
2g^{\mu \nu }\Re \left( w^{(2p+1)}_{2\mu \nu }\right) +2qA^{\mu }\Im \left( w^{(2p+1)}_{1\mu } \right) -\left[ m^{2}+\xi R + q^{2} A_{\mu }A^{\mu }  \right]  w^{(2p+1)}_{0},
\nonumber \\
\label{eq:oddw1re}
\\
0 & = & \nabla ^{\mu} \Im \left( w^{(2p+1)}_{1\mu } \right) - qA^{\mu } \nabla _{\mu }w^{(2p+1)}_{0} - q \left( \nabla ^{\mu }A_{\mu } \right) w_{0}^{(2p+1)},
\label{eq:oddw1im}
\end{eqnarray}
\end{subequations}
and
\begin{eqnarray}
0 & = & 
2\nabla ^{\mu } \Re \left( w^{(2p+1)}_{2\alpha \mu } \right)
+qA^{\mu }\nabla _{\mu }\Im \left( w^{(2p+1)}_{1\alpha }\right)
+ 2q \left( \nabla _{(\alpha }A^{\mu } \right) \Im \left( w^{(2p+1)}_{1\mu )}\right)
\nonumber \\ & & 
- \frac{1}{4} \nabla _{\alpha } \Box w^{(2p+1)}_{0}
- \frac{1}{2} R^{\mu }{}_{\alpha }w^{(2p+1)}_{0;\mu }
- \left[ \frac{1}{2}\xi R_{;\alpha } + q^{2}A^{\mu }\nabla _{\alpha }A_{\mu } \right] w^{(2p+1)}_{0}
\nonumber \\ & & 
\label{eq:oddw2}
\end{eqnarray}
respectively.

We now have the machinery required to derive expressions for the renormalized expectation values of the scalar field condensate $\langle {\hat {\Phi }}{\hat {\Phi }}^{\dagger} \rangle $, current $\langle {\hat {J}}^{\mu }\rangle$ and stress-energy tensor $\langle {\hat {T}}_{\mu \nu }\rangle $ in terms of coefficients in the expansion (\ref{eq:WTaylor}).
The renormalized expectation value of the scalar field condensate takes a simple form in terms of the regularized Green's function $G_{\rm {R}}(x,x')$ (\ref{eq:Greg})
\begin{equation}
    \langle {\hat {\Phi }}{\hat {\Phi }}^{\dagger} \rangle _{\rm {ren}}
    = \lim _{x'\rightarrow x} \Re \left\{ -{\rm {i}}G_{\rm {R}}(x,x') \right\} 
    = \alpha ^{(d)} w^{(d)}_{0}(x).
    \label{eq:condensate}
\end{equation}
The renormalized current and RSET are examined in the following subsections.

\subsection{Renormalized current}
\label{sec:RJ}

The classical current $J^{\mu }$ for a charged complex scalar field $\Phi $ is
\begin{equation}
J^{\mu } = \frac{{\rm {i}}q}{8\pi }     \left[ 
\Phi ^{*} D^{\mu }\Phi - \Phi \left( D^{\mu }\Phi \right) ^{*} 
\right]  
= -\frac{q}{4\pi } \Im \left[ \Phi ^{*}D^{\mu } \Phi  \right] .
\end{equation}
The renormalized quantum expectation value of the current is therefore given as the limit
\begin{equation}
    \langle {\hat {J}}^{\mu } \rangle _{\rm {ren}}=-\frac{q}{4\pi } \lim _{x'\rightarrow x} \Im \left\{ D^{\mu } \left[ - {\rm {i}}G_{\rm {R}}(x,x')  \right] \right\} .  
    \end{equation}
The derivative is straightforward to compute and gives
\begin{equation}
    \langle {\hat {J}}_{\mu }  \rangle _{\rm {ren}}
    = -\frac{\alpha ^{(d)}q}{4\pi } \Im \left[ D_{\mu }w^{(d)}_{0}+w^{(d)}_{1\mu }\right]
    = \frac{\alpha ^{(d)}q}{4\pi } \left\{ qA_{\mu}w^{(d)}_{0}- \Im  \left[ w^{(d)}_{1\mu}\right] \right\} .
    \label{eq:renJ}
\end{equation}
We require the renormalized expectation value of the current to be conserved, that is
\begin{equation}
    0 = \nabla ^{\mu } \langle {\hat {J}}_{\mu } \rangle _{\rm {ren}} 
    = \frac{q}{4\pi } \left\{
    q\left( \nabla ^{\mu } A_{\mu } \right) w^{(d)}_{0} + qA_{\mu }\nabla ^{\mu }w^{(d)}_{0} - \nabla ^{\mu } \Im \left[ w^{(d)}_{1\mu }\right] 
    \right\}
\end{equation}
which holds from (\ref{eq:evenw1im}, \ref{eq:oddw1im}).

\subsection{Renormalized stress-energy tensor}
\label{sec:RSET}

The classical stress-energy tensor for a massive, complex, charged scalar field $\Phi $ nonminimally coupled to the space-time curvature is
\begin{eqnarray}
T_{\mu \nu } & = & 
\frac{1}{2}\left( 1-2\xi \right)  \left[ \left( D_{\mu }\Phi \right) ^{*} D_{\nu } \Phi  +D_{\mu }\Phi \left( D_{\nu }\Phi \right) ^{*} \right] 
\nonumber \\ & & 
+\frac{1}{2} \left( 2\xi -\frac{1}{2}\right)
g_{\mu \nu } g^{\rho \sigma  }\left[ \left( D_{\rho }\Phi \right) ^{*} D_{\sigma }\Phi  + D_{\rho }\Phi \left( D_{\sigma }\Phi \right) ^{*} \right]
\nonumber \\ & & 
- \xi \left[ \Phi ^{*} D_{\mu} D_{\nu }\Phi + \Phi \left( D_{\mu }D_{\nu }\Phi \right) ^{*} \right] 
+ \xi g_{\mu \nu } \left[  \Phi ^{*}D_{\rho }D^{\rho }\Phi  + \Phi \left( D_{\rho }D^{\rho }\Phi \right) ^{*} \right]
\nonumber \\ & & 
+ \xi \left( R_{\mu \nu } - \frac{1}{2}g_{\mu \nu }R \right) \Phi ^{*}\Phi - \frac{1}{2}m^{2}g_{\mu \nu }\Phi ^{*}\Phi 
\nonumber \\ & = & 
\Re \left\{
\left( 1-2\xi \right)  \left( D_{\mu }\Phi \right) ^{*} D_{\nu } \Phi  
+ \left( 2\xi -\frac{1}{2}\right)
g_{\mu \nu } g^{\rho \sigma  } \left( D_{\rho }\Phi \right) ^{*} D_{\sigma }\Phi  
\right. \nonumber \\ & & \left.
- 2\xi \Phi ^{*} D_{\mu} D_{\nu }\Phi
+2\xi g_{\mu \nu }\Phi ^{*} D_{\rho }D^{\rho }\Phi
+ \xi \left( R_{\mu \nu } - \frac{1}{2}g_{\mu \nu }R \right) \Phi ^{*}\Phi 
\right. \nonumber \\ & & \left.
- \frac{1}{2}m^{2}g_{\mu \nu }\Phi ^{*}\Phi 
\right\} .
\label{eq:SETclassical}
\end{eqnarray}
When the scalar field is uncharged, the covariant derivatives $D_{\mu }$ in (\ref{eq:SETclassical}) are replaced by the usual space-time covariant derivatives $\nabla _{\mu }$ and the expression (\ref{eq:SETclassical}) reduces to that given in \cite{Decanini:2005eg,Fulling:1974pu,Bunch:1980vc} for a neutral scalar field.

Following the Hadamard renormalization prescription, the expectation value of the quantum stress-energy tensor operator is given as the limit
\begin{equation}
    \langle {\hat {T}}_{\mu \nu}\rangle =
    \lim _{x'\rightarrow x}\Re\left\{  {\mathcal {T}}_{\mu \nu }(x,x') \left[ - {\rm {i}}G_{\rm {R}}(x,x') \right] \right\}  ,
\end{equation}
where ${\mathcal {T}}_{\mu \nu }$ is the second order differential operator
\begin{eqnarray}
{\mathcal {T}}_{\mu \nu } & = & 
\left( 1- 2\xi  \right) g_{\nu }{}^{\nu '} D_{\mu }D^{*}_{\nu '}
+ \left( 2\xi - \frac{1}{2} \right) g_{\mu \nu }g^{\rho \tau '}D_{\rho }D^{*}_{\tau '}
-2\xi  D_{\mu }D_{\nu }
\nonumber \\ & & 
+ 2\xi g_{\mu \nu }D_{\rho }D^{\rho } 
+ \xi \left( R_{\mu \nu }- \frac{1}{2}g_{\mu \nu }R \right) - \frac{1}{2}m^{2} g_{\mu \nu },
\label{eq:RSETop}
\end{eqnarray}
where $g_{\mu }{}^{\mu '}$ is the bivector of parallel transport.
According to Wald's axioms \cite{Wald:1977up}, the renormalized expectation value of the stress-energy tensor is unique only up to the addition of a local conserved tensor. We therefore write
\begin{equation}
 \langle {\hat {T}}_{\mu \nu}\rangle _{\rm {ren}}=
\alpha ^{(d)}    \lim _{x'\rightarrow x}\Re \left[ {\mathcal {T}}_{\mu \nu }(x,x') W^{(d)}(x,x') \right] + {\widetilde {\Theta }}^{(d)}_{\mu \nu },
\end{equation}
where ${\widetilde {\Theta }}^{(d)}_{\mu \nu }$ is a local tensor whose form will be constrained by considering the divergence of $\langle {\hat {T}}_{\mu \nu}\rangle _{\rm {ren}}$.

Applying ${\mathcal {T}}_{\mu \nu }$ (\ref{eq:RSETop}) to $W(x,x')$ (\ref{eq:WTaylor}) and taking the real part, we find
\begin{eqnarray}
\langle {\hat {T}}_{\mu \nu}\rangle _{\rm {ren}}
& = & 
\alpha ^{(d)} \left\{
-2 \Re \left( w^{(d)}_{2\mu \nu }\right)
-2qA_{(\mu }\Im \left( w^{(d)}_{1\nu )}\right)
-\left( \xi - \frac{1}{2} \right) w^{(d)}_{0;\mu \nu }
\right. \nonumber \\ & & \left. 
+ \left( \xi R_{\mu \nu } + q^{2}A_{\mu }A_{\nu } \right) w^{(d)}_{0}
+g_{\mu \nu }\left[ g^{\rho \tau }\Re \left( w^{(d)}_{2\rho \tau} \right)
+ qA^{\rho }\Im \left( w^{(d)}_{1\rho }\right) 
\right. \right. \nonumber \\ & & \left. \left.
+ \left( \xi - \frac{1}{4} \right) \Box w^{(d)}_{0}
- \frac{1}{2}\left( m^{2}+ \xi R +q^{2}A_{\rho }A^{\rho } \right) w^{(d)}_{0}
\right]
 \right\} 
 \nonumber \\ & & 
  + {\widetilde {\Theta }}^{(d)}_{\mu \nu }.
  \label{eq:RSET}
\end{eqnarray}
This expression is manifestly symmetric in $\mu $ and $\nu $ and reduces to the corresponding expression in \cite{Decanini:2005eg} when $q=0$ and the scalar field is neutral.

We now examine the conservation of the RSET.
Taking the divergence of (\ref{eq:RSET}), and simplifying using the identities (\ref{eq:evenw1}, \ref{eq:evenw2}, \ref{eq:oddw1}, \ref{eq:oddw2}), we find, when $d=2p$ is even,
\begin{equation}
    \nabla ^{\mu }\langle {\hat {T}}_{\mu \nu}\rangle _{\rm {ren}} 
    = -\alpha ^{(2p)}p\nabla _{\nu }V^{(2p)}_{10} + 4\pi F_{\mu \nu }\langle {\hat {J}}^{\mu } \rangle _{\rm {ren}}  + \nabla ^{\mu }{\widetilde {\Theta }}^{(2p)}_{\mu \nu },
    \label{eq:divTeven}
\end{equation}
and when $d=2p+1$ is odd,
\begin{equation}
    \nabla ^{\mu }\langle {\hat {T}}_{\mu \nu}\rangle _{\rm {ren}} 
    =  4\pi F_{\mu \nu }\langle {\hat {J}}^{\mu } \rangle _{\rm {ren}} + \nabla ^{\mu }{\widetilde {\Theta }}^{(2p+1)}_{\mu \nu },
    \label{eq:divTodd}
\end{equation}
where we have used the expression (\ref{eq:renJ}) for the renormalized expectation value of the current.
The first term on the right-hand-side of (\ref{eq:divTeven}) arises in the neutral scalar field case \cite{Decanini:2005eg}.
Following \cite{Decanini:2005eg}, we therefore define
\begin{equation}
    {\widetilde {\Theta }}^{(d)}_{\mu \nu } = 
    \begin{cases}
    \alpha ^{(2p)}p g_{\mu \nu }V^{(2p)}_{10}+ \Theta ^{(2p)}_{\mu \nu } & d=2p,
    \\
    \Theta ^{(2p+1)}_{\mu \nu }  & d=2p+1,
    \end{cases}
    \label{eq:Thetatilde}
\end{equation}
where $\Theta ^{(d)}_{\mu \nu }$ is a local conserved tensor, giving the expected renormalization ambiguity in the RSET.
For all $d$, we then have
\begin{equation}
    \nabla ^{\mu }\langle {\hat {T}}_{\mu \nu}\rangle _{\rm {ren}} 
    =  4\pi F_{\mu \nu }\langle {\hat {J}}^{\mu } \rangle _{\rm {ren}} 
\end{equation}
leading to the nonconservation of the RSET of the charged scalar field.
This result follows from the fact that there are two matter fields in our system, the electromagnetic field (which we are treating as purely classical) and the quantum charged scalar field.
It is only the total stress-energy tensor arising from both matter fields which will be conserved.
The stress-energy tensor due to the classical electromagnetic field is
\begin{equation}
T_{\mu \nu }^{{\rm {F}}} =
   F_{\mu \rho }F_{\nu }{}^{\rho } - \frac{1}{4}g_{\mu \nu }F_{\rho \tau }F^{\rho \tau }.
   \label{eq:TEM}
\end{equation}
Taking the divergence gives
\begin{equation}
    \nabla ^{\mu }T_{\mu \nu }^{\rm {F}}
    = F_{\nu \rho} \nabla _{\mu }F^{\mu \rho} 
    = 4\pi F_{\nu \rho }\langle {\hat {J}}^{\rho } \rangle _{\rm {ren}},
\end{equation}
where we have used Maxwell's equation
\begin{equation}
    0 = \nabla _{[\mu }F_{\rho \tau ]}
\end{equation}
(which is unmodified by the presence of the charged scalar field) and the second equality follows from the semiclassical Maxwell equation (\ref{eq:SCME}).
Therefore, since the electromagnetic field $F_{\mu \nu }$ is antisymmetric, the total stress-energy tensor $T_{\mu \nu }^{\rm {F}} + \langle {\hat {T}}_{\mu \nu}\rangle _{\rm {ren}} $ is conserved, as required.

With the definition (\ref{eq:Thetatilde}), the expression (\ref{eq:RSET}) can be simplified using (\ref{eq:evenw1re}, \ref{eq:oddw1re}) to give, when $d=2p$ is even,
\begin{subequations}
\begin{eqnarray}
\langle {\hat {T}}_{\mu \nu}\rangle _{\rm {ren}} &  = & 
\alpha ^{(2p)} \left\{
-2 \Re \left( w^{(2p)}_{2\mu \nu }\right)
-2qA_{(\mu }\Im \left( w^{(2p)}_{1\nu )}\right)
-\left( \xi - \frac{1}{2} \right) w^{(2p)}_{0;\mu \nu }
\right. \nonumber \\ & & \left. 
+ \left( \xi R_{\mu \nu } + q^{2}A_{\mu }A_{\nu } \right) w^{(2p)}_{0}
+g_{\mu \nu }\left[ 
\left( \xi - \frac{1}{4} \right) \Box w^{(2p)}_{0} - V^{(2p)}_{10}
\right]
 \right\} 
 \nonumber \\ & & 
  + \Theta ^{(2p)}_{\mu \nu },
  \label{eq:RSETdeven}
\end{eqnarray}
and when $d=2p+1$ is odd
\begin{eqnarray}
\langle {\hat {T}}_{\mu \nu}\rangle _{\rm {ren}} &  = & 
\alpha ^{(2p+1)} \left\{
-2 \Re \left( w^{(2p+1)}_{2\mu \nu }\right)
-2qA_{(\mu }\Im \left( w^{(2p+1)}_{1\nu )}\right)
\right. \nonumber \\ & & \left.
-\left( \xi - \frac{1}{2} \right) w^{(2p+1)}_{0;\mu \nu }
+ \left( \xi R_{\mu \nu } + q^{2}A_{\mu }A_{\nu } \right) w^{(2p+1)}_{0}
\right. \nonumber \\ & & \left.
+g_{\mu \nu }
\left( \xi - \frac{1}{4} \right) \Box w^{(2p+1)}_{0} 
 \right\}  
 + \Theta _{\mu \nu } ^{(2p+1)} .
 \label{eq:RSETdodd}
\end{eqnarray}
\end{subequations}

\subsection{Renormalization ambiguities}
\label{sec:ambig}

The RSET constructed in the previous subsection was defined up to a local conserved tensor $\Theta ^{(d)}_{\mu \nu }$, in accordance with Wald's axioms \cite{Wald:1977up}. The possible form of $\Theta ^{(d)}_{\mu \nu }$ is discussed in detail in  \cite{Decanini:2005eg}.

There is an additional renormalization ambiguity in even space-time dimensions, due to the choice of renormalization length scale $\ell $ arising in the Hadamard parametrices (\ref{eq:GF_d2}, \ref{eq:GF_deven}).
This leads to an ambiguity in the Hadamard coefficient $W^{(2p)}(x,x')$ corresponding to the freedom to make the replacement
\begin{equation}
    W^{(2p)}(x,x') \rightarrow W^{(2p)}(x,x') + V^{(2p)}(x,x') \ln \ell ^{2}.
\end{equation}
Considering the terms in the covariant Taylor series expansion of $W^{(2p)}(x,x')$ (\ref{eq:WTaylor}), this replacement leads to
\begin{subequations}
\begin{eqnarray}
w^{(2p)}_{0} & \rightarrow & w^{(2p)}_{0}+V^{(2p)}_{00}\ln \ell ^{2},
\\
w^{(2p)}_{1\mu } & \rightarrow & w^{(2p)}_{1\mu } +V^{(2p)}_{01\mu }\ln \ell ^{2} ,
\\ 
w^{(2p)}_{2\mu \nu } & \rightarrow & w^{(2p)}_{2\mu \nu }+ \left( V^{(2p)}_{02\mu \nu }+\frac{1}{2} g_{\mu \nu }V^{(2p)}_{10} \right) \ln \ell ^{2}.
\end{eqnarray}
\end{subequations}
Ambiguities in the renormalized expectation values of the scalar field condensate, current and stress-energy tensor therefore arise.

For the scalar field condensate (\ref{eq:condensate}), we find
\begin{equation}
     \langle {\hat {\Phi }}{\hat {\Phi }}^{\dagger} \rangle _{\rm {ren}} 
   \rightarrow  
     \alpha ^{(2p)} w^{(2p)}_{0} + V^{(2p)}_{00} \ln \ell ^{2}.
\end{equation}
From (\ref{eq:V00d2}), in two dimensions $V^{(2)}_{00}$ is a constant.
In four dimensions, from (\ref{eq:V00d4}), the renormalization ambiguity depends on the Ricci scalar curvature as well as the mass and coupling of the scalar field, and vanishes when the scalar field is massless and conformally coupled.
In both two and four space-time dimensions, the renormalization ambiguity in the expectation value of the scalar field condensate does not depend on the electromagnetic potential.

The ambiguity in the renormalized expectation value of the current (\ref{eq:renJ}) is
\begin{equation}
 \langle {\hat {J}}_{\mu }  \rangle _{\rm {ren}}
    \rightarrow \frac{\alpha ^{(2p)}q}{4\pi } 
    \left\{ qA_{\mu}w^{(2p)}_{0}- \Im  \left[ w^{(2p)}_{1\mu}\right]
    +  \left( qA_{\mu } V^{(2p)}_{00} - \Im \left[ V^{(2p)}_{01\mu } \right] \right) \ln \ell ^{2} \right\} .   
\end{equation}
In two space-time dimensions, using (\ref{eq:V00d2}, \ref{eq:V01d2}), we see that the ambiguity in the renormalized expectation value of the current vanishes.
From (\ref{eq:V00d4}, \ref{eq:V01d4}), the same is not true in four space-time dimensions, when 
\begin{equation}
 \langle {\hat {J}}_{\mu }  \rangle _{\rm {ren}}
    \rightarrow \frac{\alpha ^{(4)}q}{4\pi } 
    \left\{ qA_{\mu}w^{(4)}_{0}- \Im  \left[ w^{(4)}_{1\mu}\right]
    +   \frac{q}{12}\left(\nabla ^{\rho }F_{\rho \mu } \right) \ln \ell ^{2} \right\} .  
    \label{eq:Jambig4d}
\end{equation}
Given that the current acts as a source for the semiclassical Maxwell equations (\ref{eq:SCME}), this renormalization ambiguity corresponds to a constant renormalization of the permeability of free space (which has effectively been set equal to $4\pi $ in (\ref{eq:SCME}) as we are using Gaussian units), which we discuss further in section \ref{sec:comparison}.

The renormalization ambiguity in the RSET takes the form
\begin{equation}
   \langle {\hat {T}}_{\mu \nu}\rangle _{\rm {ren}} \rightarrow 
   \langle {\hat {T}}_{\mu \nu}\rangle _{\rm {ren}} + \Psi ^{(2p)}_{\mu \nu }\ln \ell ^{2}
   \label{eq:Psidef}
\end{equation}
where the local tensor $\Psi _{\mu \nu }^{(2p)}$ is
\begin{eqnarray}
    \Psi ^{(2p)}_{\mu \nu } &  = & \alpha ^{(2p)}
    \left\{ 
    -2 \Re \left[ V^{(2p)}_{02\mu \nu } \right] 
    -2q A_{(\mu } \Im \left[ V^{(2p)}_{01\nu )} \right]
    - \left( \xi - \frac{1}{2} \right) V^{(2p)}_{00;\mu \nu }
    \right. \nonumber \\ & &  \left.
    + \left( \xi R_{\mu \nu }+q^{2}A_{\mu }A_{\nu }\right) V^{(2p)}_{00}
    +\left( \xi - \frac{1}{4} \right) g_{\mu \nu } \Box V^{(2p)}_{00}
    \right. \nonumber \\ & & \left.
    - g_{\mu \nu }V^{(2p)}_{10} \right\} .
    \label{eq:Psi}
\end{eqnarray}
In two space-time dimensions, we have, using (\ref{eq:V20expansion}, \ref{eq:V1d2}),
\begin{equation}
    \Psi ^{(2)}_{\mu \nu } = \frac{ \alpha ^{(2)} }{2} m^{2}g_{\mu \nu }.
\end{equation}
This is the same as in the neutral scalar field case \cite{Decanini:2005eg} and there are no corrections due to the electromagnetic potential.
In particular, $\Psi ^{(2)}_{\mu \nu }$ vanishes if the scalar field is massless.
For a massive scalar field, since the RSET satisfies the semiclassical Einstein equations (\ref{eq:SCEE}), the ambiguity (\ref{eq:Psidef}) corresponds to a renormalization of the cosmological constant $\Lambda $.

When $d=4$, the tensor (\ref{eq:Psi}) is much more complicated.
Using (\ref{eq:V40pexpansion}, \ref{eq:V1d4}), it is found to be
\begin{eqnarray}
    \Psi ^{(4)}_{\mu \nu } &  = &\alpha ^{(4)} \left\{  
    \frac{1}{2} \left( \xi - \frac{1}{6}\right) \left[ m^{2}+\left( \xi - \frac{1}{6} \right) R \right] R_{\mu \nu }
    + \frac{1}{120}\Box R_{\mu \nu }
    \right. \nonumber \\ & &  \left.
    - \frac{1}{2} \left( \xi ^{2}- \frac{1}{3}\xi + \frac{1}{30} \right) R_{;\mu \nu }
    - \frac{1}{90} R^{\alpha }{}_{\mu }R_{\alpha \nu }
    +\frac{1}{180}R^{\alpha \beta }R_{\alpha \mu \beta \nu }
     \right. \nonumber \\ & & \left. 
    + \frac{1}{180} R^{\alpha \beta \gamma}{}_{\mu } R_{\alpha \beta \gamma \nu }
    + \frac{q^{2}}{12}F^{\alpha }{}_{\mu }F_{\nu \alpha }
    + g_{\mu \nu }\left\{
    \frac{1}{720}R^{\alpha \beta }R_{\alpha \beta }
     \right. \right. \nonumber \\ & & \left. \left.
    - \frac{1}{720}R^{\alpha \beta \gamma \delta }R_{\alpha \beta \gamma \delta }
    + \frac{1}{2}\left( \xi ^{2} - \frac{1}{3}\xi + \frac{1}{40} \right) \Box R
    \right. \right. \nonumber \\ & & \left. \left.
    - \frac{1}{8} \left[ m^{2} + \left( \xi - \frac{1}{6}\right) R \right] ^{2} 
    + \frac{q^{2}}{48}F^{\alpha \beta }F_{\alpha \beta} 
    \right\} \right\} .
    \label{eq:Psid4}
\end{eqnarray}
In this case there are corrections arising from the electromagnetic field strength.
The curvature terms in (\ref{eq:Psid4}) correspond to higher-order corrections to the gravitational action giving rise to the semiclassical Einstein equations (\ref{eq:SCEE}). 
The corrections due to the electromagnetic field are proportional to the classical electromagnetic stress-energy tensor (\ref{eq:TEM}) and therefore the ambiguity (\ref{eq:Psidef}) in this case corresponds to a renormalization of the gravitational constant $G$ (which we have fixed by $8\pi G=1$ in (\ref{eq:SCEE})).  See section \ref{sec:comparison} for further discussion of this point.

\subsection{Trace anomaly}
\label{sec:trace}

Now suppose that the local geometric tensor $\Theta ^{(d)}_{\mu \nu }$ which arises in the RSET (\ref{eq:RSETdeven}, \ref{eq:RSETdodd}) is given by
\begin{equation}
    \Theta ^{(d)}_{\mu \nu } =
    \begin{cases}
    \Psi ^{(2p)}_{\mu \nu } \ln \ell ^{2} & d=2p, \\
    0 & d=2p+1,
    \end{cases}
\end{equation}
so that the only renormalization ambiguity we are taking into account is that due to the choice of renormalization length scale $\ell $.
We now consider the trace of the RSET.

First, from the  homogeneous scalar field equation (\ref{eq:V2eqn}, \ref{eq:V2eqndeven}) satisfied by the biscalar $V^{(d)}(x,x')$, we can establish the identity
\begin{equation}
  0  =  
2g^{\mu \nu }\Re \left( V^{(2p)}_{02\mu \nu }\right) + 2pV^{(2p)}_{10} +2qA^{\mu }\Im \left( V^{(2p)}_{01\mu } \right) -\left[ m^{2}+\xi R + q^{2} A_{\mu }A^{\mu }  \right]  V^{(2p)}_{00}  .
\label{eq:Videntity}
\end{equation}
We can then calculate the trace of $\Psi ^{(2p)}_{\mu \nu }$:
\begin{eqnarray}
    g^{\mu \nu }\Psi ^{(2p)}_{\mu \nu }
    & = & 
    \alpha ^{(2p)} \left\{
    -2g^{\mu \nu }\Re \left[ V^{(2p)}_{02\mu \nu } \right] 
    - 2q A^{\mu}\Im \left[ V^{(2p)}_{01 \mu } \right]
     - 2p V^{(2p)}_{10}
    \right. \nonumber \\ & & \left.
    + \left(2p-1 \right) \left( \xi - \frac{p-1}{2\left( 2p-1 \right)} \right) \Box V^{(2p)}_{00}
        + \left[ \xi R + q^{2}A^{\mu }A_{\mu } \right] V^{(2p)}_{00}
    \right\}
    \nonumber \\ & = & 
    -\alpha ^{(2p)}\left\{ 
    m^{2} V^{(2p)}_{00} - \left( 2p -1 \right) \left( \xi - \xi _{{\rm {c}}} \right) \Box V^{(2p)}_{00} 
    \right\} 
    \label{eq:Psitrace}
\end{eqnarray}
where $\xi _{\rm {c}}$, given by (\ref{eq:xic}), corresponds to conformal coupling and we have simplified using the identity (\ref{eq:Videntity}).
Note that the trace (\ref{eq:Psitrace}) vanishes when the scalar field is massless and conformally coupled.

The trace of the RSET (\ref{eq:RSETdeven}) is given, when $d=2p$ is even, by 
\begin{eqnarray}
  \langle {\hat {T}}_{\mu }^{\mu }\rangle _{\rm {ren}} &  = & 
\alpha ^{(2p)} \left\{
-2 g^{\mu \nu }\Re \left( w^{(2p)}_{2\mu \nu }\right)
-2qA^{\mu }\Im \left( w^{(2p)}_{1\mu }\right)
\right. \nonumber \\ & & \left.
-(2p-1)\left( \xi - \xi_{\rm {c}} \right)\Box w^{(2p)}_{0}
+ \left( \xi R + q^{2}A^{\mu }A_{\mu } \right) w^{(2p)}_{0}
\right. \nonumber \\ & &  \left.
 -2p V^{(2p)}_{10} \right\}
+ g^{\mu \nu }\Psi ^{(2p)}_{\mu \nu },   
\end{eqnarray}
which simplifies, using (\ref{eq:evenw1re}) to give
\begin{equation}
 \langle {\hat {T}}_{\mu }^{\mu }\rangle _{\rm {ren}}
 = 
 -\alpha ^{(2p)} \left\{
 m^{2} w^{(2p)}_{0} 
 -(2p-1)\left( \xi - \xi_{\rm {c}} \right)\Box w^{(2p)}_{0}
-2V^{(2p)}_{10} \right\}
+ g^{\mu \nu }\Psi ^{(2p)}_{\mu \nu }. 
\label{eq:tracedeven}
\end{equation}
A similar calculation for $d=2p+1$, using (\ref{eq:oddw1re}, \ref{eq:RSETdodd}), yields
\begin{equation}
 \langle {\hat {T}}_{\mu }^{\mu }\rangle _{\rm {ren}}
 = 
 -\alpha ^{(2p+1)} \left\{
 m^{2} w^{(2p+1)}_{0} 
 -2p\left( \xi - \xi_{\rm {c}} \right)\Box w^{(2p+1)}_{0}
 \right\} .
 \label{eq:tracedodd}
\end{equation}
If the scalar field is massless and conformally coupled, the trace (\ref{eq:tracedodd}) vanishes for odd numbers of space-time dimensions.
However, for even number of space-time dimensions, the trace does not vanish for a massless and conformally coupled scalar field.
In this case we obtain the trace anomaly
\begin{equation}
  \langle {\hat {T}}_{\mu }^{\mu }\rangle _{\rm {ren}} =2\alpha ^{(2p)}V^{(2p)}_{10}.  
\end{equation}
When $d=2$, the coupling constant for conformal coupling is $\xi _{\rm{c}}=0$, and then, using (\ref{eq:alpha2}, \ref{eq:V1d2}), the trace anomaly is
\begin{equation}
     \langle {\hat {T}}_{\mu }^{\mu }\rangle _{\rm {ren}} 
     = \frac{1}{24}R.
\end{equation}
In this case there are no corrections to the trace anomaly due to the electromagnetic field.
When $d=4$, we have $\xi _{\rm {c}}=\frac{1}{6}$ and, using (\ref{eq:alpha2p}, \ref{eq:V1d4}), the trace anomaly takes the form
\begin{equation}
   \langle {\hat {T}}_{\mu }^{\mu }\rangle _{\rm {ren}} 
   = \frac{1}{4\pi^{2} }
   \left[ 
    \frac{1}{720}  \Box R - \frac{1}{720} R^{\alpha \beta }R_{\alpha \beta }
    + \frac{1}{720} R^{\alpha \beta \gamma \delta}R_{\alpha \beta \gamma \delta } 
    - \frac{q^{2}}{48} F^{\alpha \beta }F_{\alpha \beta } 
   \right] .
   \label{eq:4dtrace}
\end{equation}
In this case we therefore have a correction due to the electromagnetic field, depending only on the field strength tensor $F_{\mu \nu }$.
This correction arises in Minkowski space-time \cite{Donoghue:2015xla}, when the curvature-dependent terms in (\ref{eq:4dtrace}) vanish.
A similar correction arises in the DeWitt-Schwinger regularization approach \cite{Herman:1995hm}, and has also been found in the context of adiabatic regularization of a charged scalar field on cosmological space-times \cite{Ferreiro:2018qdi} (although note that we use different conventions).  

\section{Comparison with other approaches to regularization}
\label{sec:comparison}

Hadamard renormalization is not the only approach to regularization of quantum fields on curved space-times. 
Other methods which have been employed for scalar fields on curved space-time backgrounds include DeWitt-Schwinger, Pauli-Villars, dimensional, zeta-function and adiabatic regularization. 
The vast majority of work in the literature concerns a neutral rather than charged scalar field. 

A notable exception to this is Boulware's early work \cite{Boulware:1978hy}, which employed the DeWitt-Schwinger method of regularization for a charged scalar field in four space-time dimensions, and demonstrated that this was essentially equivalent to Pauli-Villars and dimensional regularization.
DeWitt-Schwinger regularization in four space-time dimensions was also studied by Herman and Hiscock \cite{Herman:1995hm}, who considered Hadamard's elementary function, corresponding to the imaginary part of the Feynman Green's function. 

There is extensive work in the literature on the equivalence of various approaches to regularization for a neutral scalar field.
For example, it has been proven that Hadamard renormalization is equivalent to zeta-function regularization \cite{Moretti:1998rf,Moretti:1998rs,Hack:2012qf}.
In addition, the DeWitt-Schwinger representation of the Feynman Green's function $G^{(d)}_{\rm {F}}(x,x')$ for a neutral scalar field is a special case of the Hadamard form (\ref{eq:GF_d2}, \ref{eq:GF_deven}, \ref{eq:GF_dodd}) \cite{Decanini:2005gt}.
This means that the divergent parts of the Hadamard and DeWitt-Schwinger representations are identical and the DeWitt-Schwinger expression corresponds to a particular choice of the biscalar $W^{(d)}_{0}(x,x')$ which is undetermined in the Hadamard formalism.
Therefore the Hadamard and DeWitt-Schwinger approaches to regularization are also equivalent \cite{Hack:2012qf}.
The Hadamard and DeWitt-Schwinger representations of the Feynman Green's function depend on linear combinations of Seeley-DeWitt coefficients, which have recently been proven to be sesquisymmetric for a charged scalar field \cite{Kaminski:2019adk}.
Accordingly, one may anticipate that the analysis of \cite{Decanini:2005gt} extends to the charged case, and therefore the DeWitt-Schwinger representation of the Feynman Green's function for a charged scalar field also corresponds to a particular case of the Hadamard representation studied in this paper.

One advantage of Hadamard renormalization is that it is a very general approach to regularization, valid for any background space-time and electromagnetic potential. 
Furthermore, the results are applicable to any quantum state, providing that state is Hadamard (which is the case for physically reasonable states \cite{Fewster:2013lqa}).
The disadvantage of Hadamard renormalization is that, because it is so general, it does not explicitly give the renormalized expectation values of physical observables such as the stress-energy tensor.
In practical applications, a particular metric and electromagnetic potential will be specified, and renormalized expectation values computed on this background for one or more quantum states.
Unless there is a high degree of symmetry, it is likely that numerical calculations will be required, typically involving sums over field modes.

Friedman-Lemaitre-Robertson-Walker (FLRW) space-times, being homogeneous and isotropic, possess sufficient symmetry to be amenable to analytic calculations.
Renormalized expectation values on these backgrounds have been studied extensively within the framework of adiabatic regularization, both for neutral \cite{Fulling:1974pu,Bunch:1980vc,Parker:1974qw,Birrell:1978,Bunch:1978gb,Anderson:1987yt} and charged \cite{Ferreiro:2018qzr,Ferreiro:2018qdi} scalar fields.
The advantage of adiabatic regularization is that, in the scenarios in which it is valid, renormalized expectation values can be computed in a comparatively straightforward manner. 
The disadvantage of adiabatic regularization is that it can only be applied to those space-times, such as FLRW space-times, which have a well-defined adiabatic regime \cite{Birrell:1978}.
Furthermore, in such space-times, adiabatic regularization only applies to a single vacuum state, namely the adiabatic vacuum. 
The adiabatic vacuum has been proven to be a Hadamard state for a neutral scalar field on an FLRW space-time \cite{Junker:1996bm,Junker:2001gx}, and one would expect this to be true also in the charged case. 
Furthermore, the equivalence of adiabatic and DeWitt-Schwinger (and hence Hadamard) renormalization when the scalar field is neutral has been demonstrated explicitly via a lengthy calculation \cite{Birrell:1978,delRio:2014bpa}.
The fact that the trace anomaly (\ref{eq:4dtrace}) for a charged scalar field computed in this paper using Hadamard renormalization agrees with that obtained in \cite{Herman:1995hm} using DeWitt-Schwinger regularization and in \cite{Ferreiro:2018qdi} using adiabatic regularization provides strong evidence for the equivalence of these approaches.

Very recently, adiabatic regularization on an FLRW space-time has been employed to study the running of coupling constants in the theory considered here \cite{Ferreiro:2018oxx}.
As observed in \cite{Bunch:1979mq}, Hadamard renormalization is not the most appropriate framework for discussing the renormalization of coupling constants in the field equations, because the Hadamard parametrix depends on $\sigma ^{;\mu }$ and hence the direction in which the points are separated.  
However, in even numbers of space-time dimensions, the renormalization length scale $\ell $ which appears in the Hadamard parametrix leads to a renormalization ambiguity (as discussed in section \ref{sec:ambig}), which in turn can be interpreted as a renormalization of the coupling constants.
To see how our results compare to those derived in \cite{Ferreiro:2018oxx} using adiabatic regularization, we write the semiclassical Einstein equations (\ref{eq:SCEE}) and semiclassical Maxwell equations (\ref{eq:SCME}) with the dimensionful coupling constants restored and including the classical electromagnetic stress-energy tensor $T^{\rm {F}}_{\mu \nu }$ (\ref{eq:TEM})
\begin{subequations}
\label{eq:SCEEME}
\begin{eqnarray}
  G_{\mu \nu }+ \Lambda g_{\mu \nu } & = &  8\pi G  \left[ \langle {\hat {T}}_{\mu \nu } \rangle _{\rm {ren}} + \frac{1}{\mu _{0}} T^{\rm {F}}_{\mu \nu } \right],
  \label{eq:SCEE1}
  \\
  \nabla _{\mu } F^{\mu \nu } & = & 4\pi \mu _{0} \langle {\hat {J}}^{\nu } \rangle _{\rm {ren}} ,
  \label{eq:SCME1}
\end{eqnarray}
\end{subequations}
where $G$ is Newton's gravitational constant, $\mu _{0}$ is the permeability of free space and we have fixed the speed of light $c$ to be unity.
We consider in turn the effect of the renormalization ambiguity for the semiclassical Einstein equations (\ref{eq:SCEE1}) and Maxwell equations (\ref{eq:SCME1}) in four space-time dimensions.

The renormalization ambiguity in the RSET is given by (\ref{eq:Psidef}), where, in four space-time dimensions, the tensor $\Psi _{\mu \nu }^{(4)}$ has the form (\ref{eq:Psid4}), and consists of two types of terms.
The first type depend only on the space-time curvature and are present for a neutral scalar field with $q=0$ \cite{Decanini:2005eg}. 
In this case, as a result of renormalization, the semiclassical Einstein equations (\ref{eq:SCEE1}) are modified by the addition of local geometric tensors usually denoted by $H^{1/2}_{\mu \nu }$, which arise as functional derivatives of higher curvature terms in the renormalized effective action:
\begin{equation}
   G_{\mu \nu }+ \Lambda g_{\mu \nu } 
   + \gamma _{1} H^{1}_{\mu \nu } + \gamma _{2} H^{2}_{\mu \nu } =   8\pi G \left[ \langle {\hat {T}}_{\mu \nu } \rangle _{\rm {ren}} + T^{\rm {F}}_{\mu \nu } \right] , 
\end{equation}
where $\gamma _{1}$ and $\gamma _{2}$ are coupling constants and the geometric tensors are given by \cite{Decanini:2005eg} 
\begin{eqnarray}
  H^{1}_{\mu \nu } & = & 
  2R_{;\mu \nu } - 2 R R_{\mu \nu } + g_{\mu \nu } \left[ \frac{1}{2}R^{2} -2\Box R \right] , 
\nonumber 
  \\
  H^{2}_{\mu \nu } & = & R_{;\mu \nu } - \Box R_{\mu \nu } - 2R^{\alpha \beta }R_{\alpha \mu \beta \nu }
  + \frac{1}{2} g_{\mu \nu }\left[ R^{\alpha \beta }R_{\alpha \beta } - \Box R \right] .
\end{eqnarray}
The second type of terms in $\Psi _{\mu \nu }^{(4)}$ depend on the classical stress-energy tensor $T_{\mu \nu }^{\rm {F}}$ of the electromagnetic field.
We can write (\ref{eq:Psid4}) in the form
\begin{eqnarray}
\Psi _{\mu \nu }^{(4)} & = & \alpha ^{(4)} 
\left\{
\frac{1}{2} \left( \xi - \frac{1}{6}\right) ^{2} H^{1}_{\mu \nu }
- \frac{1}{180} \left( H^{1}_{\mu \nu } - 3H^{2}_{\mu \nu } \right)
\right. \nonumber \\ & & \left.
- m^{2} \left( \xi - \frac{1}{6} \right) \left( R_{\mu \nu } - \frac{1}{2} g_{\mu \nu }R \right)
+ \frac{1}{4}m^{2}g_{\mu \nu }
- \frac{q^{2}}{12} T_{\mu \nu }^{\rm {F}}
 \right\} .
\end{eqnarray}
Adding a term $\Psi _{\mu \nu }^{(4)} \ln \ell ^{2}$ to $\langle {\hat {T}}_{\mu \nu } \rangle _{\rm {ren}}$ (\ref{eq:Psidef}) therefore corresponds to a renormalization of the constants $G$, $\Lambda $, and $\gamma _{1/2}$ in agreement with well-known results in both adiabatic regularization \cite{Bunch:1980vc} and dimensional regularization  \cite {Bunch:1979mq} for a neutral scalar field.
The presence of the scalar field charge $q$ leads to a renormalization of the permeability of free space:
\begin{equation}
    \frac{1}{\mu _{0}} \rightarrow \frac{1}{\mu _{0}} - \frac{q^{2}\alpha ^{(4)}}{12} \ln \ell ^{2}.
    \label{eq:mu0ren}
\end{equation}

We now turn to the renormalization ambiguity (\ref{eq:Jambig4d}) in the expectation value of the current:
\begin{equation}
\langle {\hat {J}}_{\mu } \rangle _{\rm {ren}} \rightarrow \langle {\hat {J}}_{\mu } \rangle _{\rm {ren}} + \frac{\alpha ^{(4)}q^{2}}{48\pi } \left( \nabla ^{\rho }F_{\rho \mu } \right) \ln \ell ^{2}.    
\end{equation}
From the semiclassical Maxwell equations (\ref{eq:SCME1}), this yields precisely the same renormalization of the permeability of free space (\ref{eq:mu0ren}) as obtained from the semiclassical Einstein equations.
This could alternatively be interpreted as a renormalization of the scalar field charge by fixing the permeability of free space $\mu _{0}$.
This is the approach taken in \cite{Ferreiro:2018oxx} using adiabatic regularization, where it is found from the semiclassical Einstein and Maxwell equations that the scalar field charge is renormalized by terms proportional to $\ln \mu $, where $\mu $ is an arbitrary mass scale.
Therefore our results on the running of the couplings in the theory, obtained from Hadamard renormalization, are in agreement with those derived \cite{Ferreiro:2018oxx} using adiabatic regularization.

\section{Conclusions}
\label{sec:conc}

In this paper we have developed the methodology for the Hadamard renormalization of the expectation values of the scalar field condensate, current and stress-energy tensor for a massive, charged, complex scalar field with general coupling to the space-time curvature.
Our work extends the approach of \cite{Decanini:2005eg} to the charged scalar field case.
Using the Hadamard representation of the Feynman Green's function, we have 
derived the recurrence relations satisfied by the Hadamard coefficients.
Performing covariant Taylor series expansions of these coefficients, we have, in two, three and four space-time dimensions, presented sufficient terms in the expansions to enable the renormalized expectation value of the stress-energy tensor to be computed.
We have also studied the trace anomaly of the RSET, and found that while in two space-time dimensions there are no corrections due to the scalar field charge, in four space-time dimensions the trace anomaly is modified by a term depending on the electromagnetic field strength, in agreement with other approaches to renormalization \cite{Herman:1995hm,Ferreiro:2018qdi}.

The formalism developed in this paper is very general, as we make no assumptions about the background space-time metric or electromagnetic field (both of which are treated classically).  While we have presented expansions of the Hadamard coefficients explicitly for two, three and four space-time dimensions, the method can be extended to any number of space-time dimensions. 

While the methodology presented here is valid for any space-time background, we envisage that it will be particularly applicable to the calculation of the RSET for a charged scalar field on charged background black hole space-times. 
This quantity is of relevance for two physical questions.
First, the fate of the inner horizon of a charged black hole, as discussed in the introduction.
Second, the electromagnetic field of a charged black hole creates charged particle pairs which alter the Hawking emission of the black hole
\cite{Gibbons:1975kk}.
Using an adiabatic approximation in which the mass and charge of the black hole vary slowly, Hiscock and Weems \cite{Hiscock:1990ex}
considered the evolution of a charged black hole and found that it is possible for the black hole to evaporate in such a way that its charge/mass ratio increases (see also recent work by Ong and others \cite{Ong:2019rnn,Ong:2019vnv,Xu:2019wak}).
To go beyond the above adiabatic approximation, the full RSET is required.
We leave the resolution of these questions for future work.

\ack
V.B.~thanks STFC for the provision of a studentship supporting this work. 
The work of E.W.~is supported by the Lancaster-Manchester-Sheffield Consortium for Fundamental Physics under STFC grant ST/P000800/1 and partially supported by the H2020-MSCA-RISE-2017 Grant No.~FunFiCO-777740.

\end{document}